\documentstyle[12pt,epsf]{article}
\newtheorem{lem}{Lemma}
\newtheorem{prop}{Proposition}

 \font\blackboard=msbm10 
 \font\blackboards=msbm7 \font\blackboardss=msbm5
 \newfam\black \textfont\black=\blackboard
 \scriptfont\black=\blackboards \scriptscriptfont\black=\blackboardss
 \def\Bbb#1{{\fam\black\relax#1}}

\font\ninerm=cmr9
\def\uniset{\rlap{\ninerm 1}\kern.15em 1}

\def\e{\mathop{\rm e}\nolimits}
\def\i{{\rm i}}
\def\sq2{\sqrt{2}}

\title{Crystal properties of eigenstates for quantum cat maps}
\author{{\bf S. Nonnenmacher} \\
\\
CEA--Saclay, Service de Physique Th\'eorique\\
F-91191 Gif-sur-Yvette CEDEX (France)\\
{\small E-mail : nonnen@spht.saclay.cea.fr}}

\begin{document}
\maketitle
\abstract
Using the Bargmann--Husimi representation of quantum mechanics on a torus phase space, we study analytically eigenstates of quantized cat maps \cite{hannay:b}. The linearity of these maps implies a close relationship between classically invariant sublattices on the one hand, and the patterns (or `constellations') of Husimi zeros of certain quantum eigenstates on the other hand. For these states, the zero patterns are crystals on the torus. As a consequence, we can compute explicit families of eigenstates for which the zero patterns become uniformly distributed on the torus phase space in the limit $\hbar\to 0$. This result constitutes a first rigorous example of semi-classical equidistribution for Husimi zeros of eigenstates in quantized one-dimensional chaotic systems \cite{leb:vor}.

\section{Introduction}
A major question in ``quantum chaos", i.e. the study of quantum systems for which
the classical limit is chaotic, lies in the structure of eigenstates. Indeed, a `chaotic
counterpart' to the WKB-Ansatz, which holds in the case of an integrable system, is still missing (this Ansatz gives simple asymptotic formulas for individual eigenstates). The only proven results so far, collectively referred to as `Schnirelman's theorem', deal with the phase-space measures associated to eigenstates of a classically ergodic system, in the semi-classical limit. 

That is, to any quantum state $\psi_\hbar(q)$ there corresponds a phase-space measure density $H_{\psi_\hbar}(q,p)$ defined using coherent states, called the Husimi density of $\psi_\hbar$ \cite{husimi}. Now, if one quantizes an ergodic Hamiltonian system and considers a sequence of eigenstates $\{\psi_{\hbar_n}\}_{n\in{\Bbb N}}$ s.t. $\hbar_{n}\to 0$, with eigenvalues $E_n\to E$, then the associated Husimi measures $H_{\psi_n}$ will almost always converge to the Liouville measure $\mu_E$ over the energy surface $\Sigma_E$; that is, for every smooth observable $f(q,p)$, the averages $\int\!  f(q,p)\,H_{\psi_n}(q,p) dq\,dp$ tend to $\int_{\Sigma_E}\!f(q,p)\;d\mu_E$ \cite{schnir, CdV1, hellfer, zeld1}. When the ergodic classical system is not a flow, but rather a map, the same property holds for almost any semi-classical sequence of eigenstates of the quantized map, the Husimi density now converging towards the Liouville measure over the whole phase space \cite{DEGI,bouz}.

These results are very appealing in that they do not depend on the details
of the system considered, apart from its classical ergodicity. On the other hand, Schnirelman's theorem allows a minority of non-ergodic eigenstates (a set of asymptotic density zero), and we ignore what these might look like: some of them could for instance show a strong concentration along an unstable periodic
orbit (a `scar'), even in the semi-classical limit. Furthermore, the information provided by this theorem remains measure-theoretical, which is much weaker than a WKB-type Ansatz. For this reason,
we intend to study as precisely as possible the structure of eigenstates for a class of analytically tractable chaotic systems, the automorphisms of the 2-torus, or generalizations of Arnold's cat map \cite{arnold}. The quantization of these highly ergodic maps was first performed by Hannay and Berry \cite{hannay:b}, and has been rederived in different ways since then \cite{DE,bouz,zeld,klimek}; their spectral properties were analyzed by semi-classical means in \cite{keat1,keat2}, and eigenstates were studied analytically by \cite{eckhar,DE,DEGI}. 

In this paper we exhibit a certain class of quantum cat eigenstates, directly within the Bargmann--Husimi representation of quantum mechanics on the torus. This phase-space formalism, together with the linearity of the classical maps, allow us to associate a classically invariant sublattice of the unit torus with a family of eigenstates of the quantum map. Precisely, we show that the Husimi functions $H_\psi$ of these states (for which we have analytical expressions) are periodic w.r. to the invariant sublattice. We call these states crystalline, due to the patterns formed by the zeros of $H_\psi$. Moreover, we can show that in the limit of a large number of sites, these invariant sublattices (and the crystals they support) become equidistributed over the torus. Our derivation is a first step towards a still missing proof of the `uniform distribution of the Husimi zeros', conjectured by Leboeuf and Voros \cite{leb:vor,leb} for one-dimensional chaotic eigenstates. This conjecture represents a more precise statement than Schnirelman's theorem, inasmuch as the zeros completely encode the eigenstate. Unfortunately, they behave still unpredictably in the semi-classical limit, and are in general not easy to catch analytically. What saves us here is the linearity of the classical map, which yields explicit (and tractable) formulas for eigenstates. 

Before proceeding further, let us sketch the plan of the paper. After recalling the basics of quantum mechanics on the torus and describing the Bargmann and Husimi representations in this framework (section \ref{prequant}), we describe the family of dynamical systems to be studied, i.e. the quantized hyperbolic automorphisms of the 2-torus (section \ref{cat}). After explaining how to build naturally a quantum eigenstate from a classical invariant sublattice (section \ref{eigen}), we try to classify these lattices as thoroughly as possible, following the seminal work of Percival and Vivaldi \cite{percival} (section \ref{classical}). We estimate the uniformity of these invariant sets in the semi-classical limit (section \ref{asymp}), and then explicitly build different types of eigenstates (sections \ref{from}, \ref{crystal}) associated to these sets; the uniformity results can then be applied to the zero patterns of their Husimi representations.

\section{Quantum mechanics on the torus}
\label{prequant}
The classical maps we study here are defined on the unit torus phase space ${\Bbb T}^2$. Before quantizing such a map, we shall first define the quantum kinematics, i.e. the Hilbert space(s) of quantum states corresponding to this classical phase space. We will not resort to geometric
quantization, as was done in \cite{DB}, but use instead a more pedestrian approach, which yields basically the same results (it amounts to a choice of 
polarized sections parallel to the $p$ axis). 
Namely, we build our quantum states by imposing the following quasi-periodicity conditions upon a one-dimensional wave-function $|\psi\rangle$, i.e. a distribution in $\cal{S}'(\Bbb{R})$ \cite{leb:vor,bouz}:
\begin{eqnarray}
\langle q+1|\psi\rangle &=& \e^{2\i\pi\kappa_1}\,\langle q|\psi\rangle\\
\langle p+1|\psi\rangle &=& \e^{-2\i\pi\kappa_2}\,\langle p|\psi\rangle\nonumber
\end{eqnarray}
with usual Dirac notations for position and momentum representations. The Floquet parameter $\kappa=\kappa_1 +\i\kappa_2$ can take any value on the dual torus. These conditions force Planck's constant to take discrete values $\hbar=(2\pi N)^{-1}$, with $N$ a positive integer. For given $N$ and $\kappa$, the quasi-periodic states form a $N$-dimensional Hilbert space of distributions ${\cal H}_{N,\kappa}$, having for example the following basis:
\begin{equation}
\label{qj}
|q_j\rangle_{N,\kappa} = \sum_{m\in\Bbb Z} \e^{2\i\pi m\kappa_1}|q_j+m\rangle, \quad\mbox{for}\ j=0,\cdots,N-1
\end{equation}
where the allowed position peaks are discrete, $q_j=2\pi\hbar(\kappa_2+j)$. The Hermitian structure on ${\cal H}_{N,\kappa}$ is simply given by $\langle q_k|q_j\rangle_{N,\kappa}=\delta_{jk}$, for $j,k=0,\cdots,N-1$. Through a finite Fourier transform of $|q_j\rangle_{N,\kappa}$, one obtains the basis built by periodization of momentum eigenstates, $|p_m\rangle_{N,\kappa}$, with the dual discretization $p_m=2\pi\hbar(\kappa_1+m)$. 

It can be helpful to work in a phase-space representation, since we are interested in {\it semi-classical} features of states in ${\cal H}_{N,\kappa}$ (which arise for large values of $N$). We will use the Bargmann representation of quantum states on the real line \cite{barg}, simply extending it to the above distribution spaces by linearity \cite{leb:vor}. This transformation maps a Schr\"odinger wave-function $\langle q|\psi\rangle$ into an entire function
$\langle z|\psi\rangle$, through the integral kernel
\begin{equation}
\label{coh}
\langle z|q\rangle = {1\over(\pi\hbar)^{1/4}}\,\e^{-{1\over 2\hbar}(z^2+q^2)+{\sqrt{2}\over \hbar}zq},
\end{equation}
where $|z\rangle$ is a (Weyl) coherent-state, which is localized at the phase-space point $(q_o,p_o)$ given by $z={q_o-\i p_o\over\sqrt 2}$. Although $|z\rangle$ depends on $\hbar$, this dependence will not be explicit in our subsequent notations : the bra $\langle z|$ will always correspond to the value of $\hbar$ ($=1/2\pi N$) labelling the attached ket $|\psi\rangle_{N,\kappa}$.

 The combination of the equations (\ref{qj},\ref{coh}) leads to the following representation of the basis $|q_j\rangle_{N,\kappa}$ :
\begin{equation}
\label{basis}
\langle z|q_j\rangle_{N,\kappa} ={1\over(\pi\hbar)^{1/4}} \e^{2\pi N(-{1\over 2}(z^2+q_j^2)+\sqrt{2}zq_j)}\;\theta_3\left( \i\pi N(q_j-\i{\kappa_1\over N}-\sqrt{2} z)\Bigl|\i N\Bigr.\right)
\end{equation}
where we used Whittaker and Watson's \cite{WW} definition for the Jacobi theta function,
\begin{equation}
\label{thetaWW}
\theta_3(Z|\tau)=\sum_{n\in\Bbb Z} \e^{\i\pi\tau n^2 + 2\i nZ}, \quad\mbox{for  } Z,\tau\in\Bbb C,\ \Im(\tau) >0.
\end{equation}
It can be convenient to use a more general definition of theta function \cite{KP}:
\begin{eqnarray}
\Theta_{N,q_j}(\tau, Z,t) &=& \e^{-2\i\pi Nt}\sum_{\gamma \in q_j+\Bbb Z}\e^{\i\pi N\tau \gamma^2 - 2\i\pi N\gamma Z}\\
&=&\e^{-2\i\pi Nt}\;\e^{-2\i\pi NZq_j}\;\e^{\i\pi N\tau q_j^2}\;\theta_3(-\pi N(Z-\tau q_j)|N\tau),\nonumber
\end{eqnarray}
in which the basis vectors of ${\cal H}_{N,\kappa}$ read as
\begin{equation}
\langle z|q_j\rangle_{N,\kappa} ={1\over(\pi\hbar)^{1/4}}\;\Theta_{N,q_j}\!\!\left(\i,\; \i\sq2 z-{\kappa_1\over N},\;q_j{\kappa_1\over N}-\i{z^2\over 2}\right).
\end{equation}
We will use the latter type of theta functions when studying the action of the symplectic group upon biperiodic states, in the next section. On the other hand, the analytic properties of the ``old'' theta function $\theta_3$ are thoroughly described in \cite{WW}, and we need them too. 

In the Bargmann representation, the vectors of ${\cal H}_{N,\kappa}$ are completely characterized by the following quasi-periodicity properties \cite{leb:vor}:
\begin{eqnarray}
\langle z+1/\sq2 |\psi\rangle_{N,\kappa} &=& \e^{2\i\pi\kappa_1}\e^{\pi N(1/2+\sq2 z)}\ \langle z|\psi\rangle_{N,\kappa}\\
\langle z+\i/\sq2 |\psi\rangle_{N,\kappa} &=& \e^{2\i\pi\kappa_2}\e^{\pi N(1/2-\i\sq2 z)}\ \langle z|\psi\rangle_{N,\kappa}.\nonumber
\end{eqnarray}
The functions $\langle z|\psi\rangle_{N,\kappa}$ are thus quasi-periodic w.r. to the torus $T_{\Bbb C}$ of periods $1/\sq2,\i/\sq2$; they actually represent particular holomorphic sections of a complex line bundle over this torus \cite{KP, klimek, zeld}. 

The scalar product of two vectors in ${\cal H}_{N,\kappa}$ is easily computed in Bargmann's representation, thanks to a closure formula:
\begin{equation}
\label{closure}
\langle \psi|\psi'\rangle_{N,\kappa} = 2\int_{T_{\Bbb C}} d\Re z\,d\Im z\ 
\overline{\langle z|\psi\rangle}_{N,\kappa}\langle z|\psi'\rangle_{N,\kappa}\;\e^{-2\pi N|z|^2}. 
\end{equation}

Meanwhile, the positions of the zeros of a function $\langle z|\psi\rangle_{N,\kappa}$ are well-defined on $T_{\Bbb C}$; there are exactly $N$ of them, and their sum is constrained by 
\begin{equation}
\label{sum}
 \sq2\ \sum_{k=1}^N z_k =  N\left({1+\i\over 2}\right) - \i\kappa \ \bmod  [1,\i].
\end{equation}
The knowledge of the zeros allows to rebuild the quantum state, up to a constant factor $C$, from the unique function $\chi(z)=\langle z|q=0\rangle_{N=1,\kappa=0}$ : if we know the zeros $z_k$ of a Bargmann wavefunction $\langle z|\psi\rangle_{N,\kappa}$ (which satisfy (\ref{sum})), we recover this function multiplicatively:
\begin{equation}
\label{fact}
\langle z|\psi\rangle_{N,\kappa} = C\;\exp\left(2\pi z\sum_{k=1}^N (\bar{z}_k-{1-\i\over 2\sq2})\right)\,\prod_{k=1}^N \chi\left(z+{1+\i\over 2\sq2}-z_k\right).
\end{equation}
(The whole product is independent of the determination of the zeros we have used to build it.) This factorization property makes the Bargmann zeros a particularly interesting tool to construct or characterize a quantum state \cite{leb,leb:vor}, especially in the semi-classical framework. On the one hand, the set of zeros forms an {\it exact} and {\it minimal} representation of the quantum state, on the other hand the zeros live in the classical phase space, as opposed to the usual Schr\"odinger coefficients. This is the reason why we intend to investigate their positions for eigenstates of quantized cat maps. This task is a priori not trivial because the quantum state depends non-linearly of its zeros (the difficulty is similar to that of relating the coefficients and the roots of a polynomial). 

We are however able to give a complete description of the Bargmann zeros for particular states, namely the basis states of ${\cal H}_{N,\kappa}$ (\ref{basis}), since we know where $\theta_3(\pi Z|\tau)$ vanishes \cite{WW} : it has a unique zero in the fundamental torus of modulus $\tau$, lying at $Z=(1+\tau)/2$. Therefore, the $N$ zeros of $\langle z|q_j\rangle_{N,\kappa}$ have the positions
\begin{eqnarray}
\label{zeros}
\sq2 z_k &=& q_j +{1\over 2}- \i{\kappa_1\over N} +\i{(k+1/2)\over N}\, \mbox{ mod } [1,\i],\ k=0,\cdots, N-1\\
         &=& {1\over N}(j-\i\kappa +{\i\over 2}) + {1\over 2}+\i{k\over N}\,\mbox{ mod } [1,\i],\nonumber
\end{eqnarray}
 i.e., they lie along a vertical line on the torus, at maximal distance from the line $\{q=q_j\}$ (see figure 1). Actually, the Bargmann function $\langle z|q_j\rangle_{N,\kappa}$ verifies the `fine' quasi-periodicity property
\begin{equation}
\label{fine}
\langle z+\i /N\sqrt{2}|q_j\rangle_{N,\kappa}= \e^{2\i\pi q_j}\e^{\pi (1/2N-\i\sq2 z)}\ \langle z|q_j\rangle_{N,\kappa}\ .
\end{equation}

\bigskip
To close this section, let us describe the phase-space measure we can build from the Bargmann representation, i.e. the Husimi measure density \cite{husimi}. For any state $|\psi\rangle_\hbar$ in $\cal{S}'(\Bbb R)$, it is defined by the positive density
\begin{equation}
H_\psi(z,\bar z)=|\langle z|\psi\rangle|^2 \,\e^{-{z\bar z\over\hbar}}.
\end{equation}
For a state $|\psi\rangle_{N,\kappa}$, the Husimi density function is single-valued on $T_{\Bbb C}$. Like the Bargmann function, it factorizes
easily in terms of its zeros $z_k$ and of the Husimi density of $\chi(z)$:
\begin{equation}
H_\psi(z,\bar z)=\,K\,\prod_{k=1}^N\,H_\chi(z+{1+\i\over 2\sq2}-z_k).
\end{equation}
where $K=|C|^2$ is the constant which makes $H_\psi(z,\bar z)$ the density of a probability measure on $T_{\Bbb C}$, as long as $|\psi\rangle_{N,\kappa}$ is normalized (cf. eq. \ref{closure}).

\paragraph{What is known about the Husimi density ?}

The factorization properties of Bargmann and Husimi functions mean that all the information about them (except an arbitrary factor) is contained in the positions of their zeros. We now roughly sketch the relation
between the global features of the Husimi densities on the one hand, the patterns of zeros on the other hand, for two different types of eigenstates. 

For an integrable system (e.g. a time-independent Hamiltonian system on the torus), the Husimi measures of a sequence of eigenstates $\{\langle z|\psi\rangle_{N,\kappa}\}_{N\in\Bbb N}$ with energies $E_N\to E$ are known to concentrate along the orbit of energy $E$ \cite{taka}. In the same time, the zeros of these Husimi densities line up along certain classically defined lines, far away from the classical orbit (namely, anti-Stokes lines : see for instance \cite{vor89}). These results are mere consequences of the WKB asymptotic form of the eigenstates of integrable systems. As an example, one can consider the position eigenstates $|q_j\rangle_{N,\kappa}$ (see figure 1); for any $Q\in[0,1]$, we can select a sequence of states $\{|q_{j(N)}\rangle_{N,\kappa}\}_{N\in \Bbb N}$, such that $q_{j(N)}\to Q$ as $N\to\infty$. The zeros of $\langle z|q_{j(N)}\rangle_{N,\kappa}$ lie on vertical lines which converge to the line $\{q=Q+1/2\}$, whereas the Husimi densities concentrate semi-classically upon the line $\{q=Q\}$.

\medskip
On the opposite, in the case of classically ergodic dynamics (e.g. some kicked systems, or the quantum cat maps we will study here), the only analytical result concerning the eigenstates is provided by Schnirelman's theorem, described in the introduction. For almost every sequence of eigenstates $\{\psi_{N}\}_{N\in{\Bbb N}}$ of the corresponding quantum maps, the Husimi measures converge to the Liouville measure on ${\Bbb T}^2$ in the weak-$*$ sense, which means that for any continuous function $f$ on the torus, $\int_{{\Bbb T}^2} f(q,p) H_{\psi_N}(q,p)\,dq\,dp$  converges to the ergodic average $\int_{{\Bbb T}^2} f(q,p)\,dq\,dp$ as $N\to\infty$ \cite{DEGI,bouz} (the interested reader can find precisions about the weak-$*$ topology in \cite{reed}). 

This weak-$*$ convergence does not provide very precise information about the eigenstates. Moreover, it would be convenient to obtain a direct description of the phase-space minimal data formed by the zeros. Numerical calculations have yielded the following conjecture \cite{leb:vor,leb}: one expects the zeros of a
chaotic eigenstate to {\it spread uniformly} over the whole phase space (i.e. the torus) \cite{leb:vor,leb}; further studies showed that the zeros locally behave as roots of random polynomials \cite{leb:shukla} (in particular, they tend to repel each other at short distance). This uniformity would be, in a certain sense, dual to the uniformity of the Husimi measure itself. In the following, we will be able to prove such a semi-classical equidistribution for very particular cat eigenstates, in which case the zero patterns are crystalline.  

\section{Classical and quantum cat maps}
\label{cat}
The classical transformations we consider are the hyperbolic automorphisms of the unit torus phase space $\Bbb T^2$, given by matrices $S=\left(\begin{array}{cc} a&b\\ c&d \end{array}\right)$ in ${\rm SL}(2,\Bbb Z)$ with tr$(S)=a+d>2$ (this has the useful consequence $bc\neq 0$). Such a matrix acts linearly on the torus:
\begin{equation}
\vec{x}=\left(\begin{array}{c} q\\p\end{array}\right)\mbox{ mod } 1\longrightarrow S\,\vec{x}\ \mbox{ mod } 1.
\end{equation} 
These diffeomorphisms of ${\Bbb T}^2$ are known to be fully ergodic and mixing \cite{arnold}; in particular, for a sufficiently smooth observable on the torus $f(\vec{x})$, we have the property:
\begin{equation}
\label{ergodic}
\mbox{ for almost all }\vec{x}\in \Bbb T^2,\ \ \lim_{M\to\infty}{1\over M}\sum_{n=1}^M f(S^n \vec{x})\,=\,\int_{\Bbb T^2} f(\vec{x}) d^2 x,
\end{equation}
where $d^2 x$ is the usual Liouville measure on $\Bbb T^2$. The condition ``for a.a. $\vec{x}$" is not gratuitous: in the present case, all points on the torus with rational coordinates are periodic under the action of $S$, so the ergodic property (\ref{ergodic}) does not hold for them. The periodic orbit structure of classical cat maps was largely studied in \cite{percival, bart, keat1}, and we will use some of those results in the following. 

The quantization of $S$ can be performed in two steps \cite{keat2}. First, any transformation $S$ in ${\rm SL}(2,\Bbb R)$ acting  on the whole plane can be considered as the time-one map of a certain quadratic Hamiltonian flow, as long as tr$(S)\geq -2$. For arbitrary $\hbar$, evolving the Weyl-quantized Hamiltonian during a unit time yields the unitary operator $U_S$ on $L_2(\Bbb R)$, given by the integral kernel
\begin{equation}
\label{kernel}
\langle q'|U_S|q\rangle = {1\over\sqrt{2\i\pi\hbar b}} \e^{ {\i\over \hbar}[(dq'^2-2q' q+aq^2)/2b]}
\end{equation}
(This corresponds exactly to the Van Vleck quantization scheme, since the term between brackets is the generating function of the symplectic transformation $S$). The ambiguity of the square-root sign is natural: the quantization $S\longrightarrow U_S$ only provides a representation of ${\rm SL}(2,\Bbb R)$ up to a sign, called the metaplectic representation ${\rm Mp}(2,\Bbb R)$. Elements of the group ${\rm Mp}(2,\Bbb R)$ are precisely specified by both the symplectic matrix $S$ they are built from, and a holomorphic function on the upper half-plane $j(\tau)$ s.t. $j^2(\tau)=b\tau + d$. This function appears naturally if we write the above kernel in a mixed Bargmann--Schr\"odinger representation:
\begin{equation}
\langle z|U_S|q\rangle =  {1\over(\pi\hbar)^{1/4}}{1\over j(\i )} \exp\left\{ {1\over 2\hbar(\i b+d)} \left( -(d-\i b)z^2-(a-\i c)q^2+2\sq2 zq\right)\right\}.
\end{equation}
This formula is valid for any value of $\hbar$. However, in the following, working on the torus imposes the values $\hbar=(2\pi N)^{-1}$.

Now that we have quantized the symplectic transformation $S\in {\rm SL}(2,\Bbb R)$ into a unitary  operator $U_S$ on $L_2(\Bbb R)$, we extend the action of $U_S$ to the distribution spaces ${\cal H}_{N,\kappa}$ by linearity (see Appendix A). Using the notation $S(\tau) = {a\tau+c\over b\tau+d}$, we obtain
\begin{eqnarray}
\label{transf}
\langle z|U_S|q_k\rangle_{N,\kappa} &= &\langle z|U_S|q_k\rangle \;\theta_3\left(\pi N[ S(\i ) q_k -\i\sq2{z\over \i b+d} +{\kappa_1\over N}]\Bigl| S(\i )\,N\Bigr.\right)\\
&=&{1\over(\pi\hbar)^{1/4}}\,{1\over j(\i )}\;\Theta_{N,q_k}\!\!\left( S(\i ), {\i\sqrt{2}z\over \i b+d}-{\kappa_1\over N},q_k{\kappa_1\over N}-\i{z^2\over 2}-{bz^2\over \i b+d}\right).\nonumber
\end{eqnarray}
This represents a generalization (to all $\kappa$) of the action of ${\rm Mp}(2,\Bbb Z)$ on theta functions, described for instance in \cite{KP}. More precisely, the transformation $U_S$ acts onto the function $\Theta_{N,q_k}\!(\tau,Z-\kappa_1/N,t)$ according to:
\begin{equation}
\label{linear}
\Theta_{N,q_k}\!(\tau,Z-{\kappa_1\over N},t)\stackrel{U_S}{\longrightarrow}\  {1\over j(\tau)}\ 
\Theta_{N,q_k}\!\left( S(\tau),{Z\over b\tau+d}-{\kappa_1\over N},t+{bZ^2\over 2(b\tau+d)}\right).
\end{equation}
 To obtain (\ref{transf}), i.e. the transformation rule for the basis states of ${\cal H}_{N,\kappa}$, we only need to select the three parameters:  $\tau=\i,Z=\i\sq2 z,t=-\i z^2/2+q_k\kappa_1/N$.   

\medskip
For an arbitrary transformation $S\in {\rm SL}(2,\Bbb R)$, the transformed Bargmann function is not an element of ${\cal H}_{N,\kappa}$; it is a theta function, quasi-periodic w.r. to another lattice:
\begin{eqnarray}
\label{period}
\langle z+{a-\i c\over\sq2}|U_S|\psi\rangle_{N,\kappa} &=&  \e^{2\i\pi\kappa_1}\,\e^{\pi N[{a^2+c^2\over 2}+\sq2 z(a+\i c)]}\;\langle z|U_S|\psi\rangle_{N,\kappa}\\
\langle z+{-b+\i d\over\sq2}|U_S|\psi\rangle_{N,\kappa} &=& \e^{2\i\pi\kappa_2}\,\e^{\pi N[{b^2+d^2\over 2}+\sq2 z(-b-\i d)]}\;\langle z|U_S|\psi\rangle_{N,\kappa}.\nonumber
\end{eqnarray}
If $S$ is a matrix with integer entries, this lattice coincides with the period lattice of $T_{\Bbb C}$ by modularity. The space ${\cal H}_{N,\kappa}$ is then mapped by $U_S$ into ${\cal H}_{N,\kappa'}$, with \cite{DEGI,bouz}
\begin{equation}
\label{angle}
\left(\begin{array}{c}\kappa_2'\\\kappa_1'\end{array}\right) = \,S\, \left(\begin{array}{c}\kappa_2\\\kappa_1\end{array}\right) \,+\,{1\over 2}\left(\begin{array}{c} Nab\\Ncd\end{array}\right)\mbox{ mod }1.
\end{equation}
For simplicity, we will generally use the same notation $U_S$ for the unitary
operator on $L_2(\Bbb R)$ and the unitary $N\times N$ matrices connecting ${\cal H}_{N,\kappa}$ and ${\cal H}_{N,\kappa'}$, since the latter are extensions of the former to some distribution spaces.
In order to avoid the inhomogeneous term in eq. (\ref{angle}), we choose (cf.\cite{hannay:b, DEGI}) to restrict ourselves to transformations $S$ of the following types:
\begin{equation}
\label{subgroup}
\left(\begin{array}{cc} odd&even\\even&odd\end{array}\right)\quad\mbox{or}\quad\left(\begin{array}{cc}even&odd\\odd&even\end{array}\right).
\end{equation}
These transformations form a subgroup of index $3$ in ${\rm SL}(2,\Bbb Z)$, named $\Gamma_{1,2}$ in the literature \cite{mumford}. For all values of $N$, the spaces ${\cal H}_{N,\kappa}$ for $\kappa=0$ (periodic wavefunctions) and
$\kappa=\hat{0}=(1+\i)/2$ (anti-periodic wavefunctions) are both invariant under all operators $U_S$, $S\in \Gamma_{1,2}$. We thus obtain two families of (non-faithful) finite-dimensional representations of this group. On the other
hand, according to equation (\ref{angle}) an individual operator $U_S$ can act as an endomorphism on ${\cal H}_{N,\kappa}$ for some other values of $\kappa$ as well (see Appendix A, eq. (\ref{endom})).

\section{Strategy: from zeros to eigenstates}
\label{eigen}
 From now on, we give ourselves a fixed hyperbolic transformation $S\in \Gamma_{1,2}$, and study the eigenstates of the corresponding matrices $U_S$ on the above-mentioned invariant Hilbert spaces, for any value $N$. 

Analytical expressions have already been obtained for eigenstates of $U_S$ on ${\cal H}_{N,0}$ \cite{eckhar,keat2,DEGI}. The following trick was used: for all $N$, there exists an integer period $p(N)$, s.t. $U_S^{p(N)}=\e^{2\i\pi\sigma(N)}\,\uniset$, where $\uniset$ is the identity over ${\cal H}_{N,0}$, and $\sigma(N)$ is a phase. Then, starting from any state $|\psi\rangle_{N,0}$, suitable combinations of its images  $U_S^k|\psi\rangle_{N,0}$ for $\ k=1,\cdots, p(N)$ yield an orthogonal family of eigenvectors spanning a subspace of ${\cal H}_{N,0}$, with eigenangles regularly spaced on the unit circle. One can then repeat this construction from a new vector $|\psi'\rangle_{N,0}$ orthogonal to  this subspace and so on, until a complete basis is formed. Actually, we will use a similar approach in section (\ref{crystal}). 

Unfortunately, the resulting formulas for the eigenstates are not very suggestive of their "shape", in particular the limit $N\to\infty$ looks quite untractable. By contrast, the main task of this article is to construct subfamilies of eigenstates characterized by interesting phase-space features, namely crystalline structures, when expressed in the Bargmann representation.
Our approach will be similar to the one developed by Degli Esposti, Graffi and Isola \cite{DEGI}: we will assign to $N$ certain values, selected according to their arithmetical properties (we generalize the prime values considered in \cite{DEGI}), and for these $N$ we will exhibit some eigenstates showing lattice-like, or more generally crystalline zero patterns. We will use the
close link between quantum and classical motions, due to the linearity of the classical map. Indeed, our main tool is the following statement :

\begin{prop}

The action of $U_S$ on any basis state $|q_j\rangle_{N,\kappa}$ transforms the zeros of its Bargmann function $\langle z|q_j\rangle_{N,\kappa}$ according to the classical map $S$ (up to complex conjugation).
\end{prop}
 This statement follows directly from eq. (\ref{transf}), since we know where the function $\theta_3$ vanishes . Calling the transformed zeros $z_k'$, let us relate the positions of $z_k$ (eq. \ref{zeros}) and $z_k'$ :
\begin{eqnarray}
\label{zerotrans}
\sq2\ z_k&=& \left(q_j+1/2\right) + \i{1\over N}\left(1/2+k-\kappa_1\right)\quad\bmod [1,\i ]\\
\sq2\ z_k'&=& (a-\i c)\left(q_j+1/2\right)\,+\,(-b+\i d){1\over N}\left(1/2+k-\kappa_1\right)\quad\mbox{mod }[1,\i ].\nonumber
\end{eqnarray}
This can be written, with obvious notations :
\begin{eqnarray}
\label{clas}
z_k'&=&\left(\begin{array}{cc}a&-b\\-c&d\end{array}\right)z_k\quad\bmod [1/\sq2,\i/\sq2 ]\\
\Longleftrightarrow\qquad\bar{z}_k'&=&\,S\,\bar{z}_k\quad\bmod [1/\sq2,\i/\sq2]\nonumber
\end{eqnarray}
(the complex conjugation is due to the convention $z={q-\i p\over \sq2})$. 

The above lemma can be extended to {\it the images of the basis states $|q_j\rangle_{N,\kappa}$ under} ${\rm Mp}(2,\Bbb Z)$, thanks to group properties. Indeed, if we note $z_k, z_k', z_k''$ the zeros of $\langle z|q_j\rangle_{N,\kappa},\,\langle z|U_{S'}|q_j\rangle_{N,\kappa},\,\langle z|U_{S''}U_{S'}|q_j\rangle_{N,\kappa}$
respectively (indexing the zeros correctly), the group law $U_{S''}U_{S'}=\pm U_{S''S'}$ combined with eq.(\ref{clas}) imply
\begin{equation}
\bar{z}_k''=S''\,S'\bar{z}_k=S''\bar{z}_k',
\end{equation}
which shows that the zeros of $\langle z|U_{S'}|q_j\rangle_{N,\kappa}$ move classically under $U_{S''}$.  

We stress that the transformation law (\ref{clas}) is only valid for {\it very special} states, and is not at all generic in ${\cal H}_{N,\kappa}$ (In section (\ref{still}), we build another type of states for which (\ref{clas}) also applies, but this construction is possible only when $N$ is a perfect square).  

In the following, any state $|\psi\rangle$ obtained as the image of a basis state $|q_j\rangle_{N,\kappa}$ under an operator $U_S$ will be called a ``pure theta state", since its Bargmann function can be written as a single theta function times an exponential factor (eq. \ref{transf}). According to eq.(\ref{zerotrans}), the $N$ zeros of such a state make up a discrete `string' on the torus $T_{\Bbb C}$, of the form
\begin{equation}
\label{line}
z'_k=z'_o+k(n+\i m)/N\sq2,\quad \mbox{for}\quad k=0,\cdots, N-1.
\end{equation}
 Conversely, a state for which the zero pattern takes the form (\ref{line}), with coprime integers $n,m$, {\it is} a pure theta state in the space ${\cal H}_{N,\kappa}$ ($\kappa$ is drawn from the sum rule (\ref{sum})). For each $N$, we thus have a characterization of the set of pure theta states $\{U_S|q_j\rangle_{N,\kappa},\;S\in {\rm SL}(2,{\Bbb Z}),\,\kappa\in [0,1[^2,\,j=0,\ldots, N-1\}$ through their Bargmann zeros.

 Performing a dilation of factor $N\sq2$ and a reflection w.r. to the real axis, we map the lattice of zeros into an affine lattice $\Lambda=\vec{v}_o+\Lambda_{\rm lin}$, where $\Lambda_{\rm lin}$ is the integer linear lattice generated by $\vec{v}=\left(\begin{array}{c} n\\-m \end{array}\right)$ modulo the square torus of side $N$. From the above properties, we deduce the following key property:

\begin{prop}

If there exists a pure theta state $|\psi\rangle_{N,\kappa}$ whose corresponding affine string $\Lambda$ is invariant under $S$, then this state is an eigenstate of $U_S$ (and automatically belongs to one of the spaces ${\cal H}_{N,\kappa}$ on which $U_S$ acts as an endomorphism). Besides, the invariance of $\Lambda$ implies that of $\Lambda_{\rm lin}$ by linearity.
\end{prop}

In mathematics (see for instance \cite{lang}, p.81--84), the two-dimensional lattice of integer points modulo $N$ constitutes the {\it module} ${\Bbb Z}_N^2$ (here ${\Bbb Z}_N$ denotes the ring of integers 
modulo $N$; in this context, the integer $N$ is called a {\it modulus} \cite{percival}). The sublattice $\Lambda_{\rm lin}$ is a {\it submodule of type $1$} (because it is generated by $\vec v$ alone in ${\Bbb Z}_N$), or {\it principal submodule}, which is {\it free} if $n,m,N$ are globally coprime, or, equivalently, if $\Lambda_{\rm lin}$ contains $N$ points (by definition, a module is free if its generators are linearly independant). The submodules of ${\Bbb Z}_N^2$ we will consider will be principal, unless stated otherwise. 

In view of the above proposition, our first task will thus consist in identifying the integer strings $\Lambda_{\rm lin}$ invariant modulo $N$, which actually amounts to diagonalizing $S$ over ${\Bbb Z}_N^2$. The next 
section deals with the classification of these free principal eigenmodules, for a given 
transformation $S\in\Gamma_{1,2}$ and a general value $N$. In a following stage, we will go back to quantum mechanics, giving explicit formulas for
the eigenstates corresponding to the invariant strings $\Lambda$ translated from $\Lambda_{\rm lin}$. Finally, to obtain crystalline
eigenstates, we will use sublattices also built through translations of $\Lambda_{\rm lin}$, but not invariant under $S$.

\section{Free eigenmodules}
\label{classical}
Given a fixed transformation $S$, we classify its free eigenmodules $\Lambda_{\rm lin}$ for a general value of $N$ (in this section, we will drop the subscript $_{\rm lin}$ and denote a linear string $\Lambda$). This classification deeply relies on the arithmetical properties of $N$. The values $N$ for which there exists at least one free eigenmodule are called {\it admissible} (this property obviously depends on the matrix $S$). We use here a different formalism from Percival and Vivaldi's \cite{percival}, but our results highly overlap
with theirs (we do not need the complete classification of periodic orbits these authors have worked out).

When diagonalizing $S$ as a hyperbolic matrix acting on ${\Bbb R}^2$, we obtain the following formulas for eigen-values and -vectors : 
\begin{equation}
\label{lambda}
\lambda_\pm={a+d\pm\sqrt{(a+d)^2-4}\over 2}\mbox{  is associated to the eigenvector  }\left(\begin{array}{c} 1\\k_\pm={\lambda_\pm-a\over b}\end{array}\right).
\end{equation}
The important quantity here is the argument under the square-root, i.e. the discriminant of $S$. We choose to normalize this discriminant as $D=({a+d\over 2})^2-1$ in the following ($D$ is integer according to eq.(\ref{subgroup})). For the case of hyperbolic maps we consider ($|a+d|>2$), the square-root of $D$ in $\Bbb R$ is always irrational, so there is no eigenvector of $S$ on $\Bbb R^2$ with rational coordinates. 

On the other hand, $S$ can have integer eigenvectors when acting on ${\Bbb Z}_N^2$, provided the discriminant $D$ admits an integer square root {\it modulo N}. This property is an arithmetical one, and we will need to factorize $N$ as a product of {\it primary} numbers (i.e. powers of a prime). 

Before proceeding further, we fix our arithmetical notations: if $M$ is a divisor of $N$, we note $M|N$; the greatest common divisor (gcd) of a family of integers will be noted $(n_1\cap n_2\cap,\cdots,\cap n_r)$, and $N_l\stackrel{\rm def}{=}N/(N\cap l)$ for any $l\in{\Bbb Z}_N$. The simplest case to deal with is $N=P$ a
prime number; $\Bbb Z_P$ is then a field, and $\Bbb Z_P^2$ a vector space. The existence problem of a square-root of $D$ modulo $P$ yields a partition of the primes $P\geq 3$ \cite{serre,hardy}: if there exists an integer $x$ {\it coprime to} $P$ s.t. $x^2\equiv D$ mod $P$, $D$ is said to be a quadratic residue of $P$, or equivalently, $P$ is said to be {\it splitting} w.r.to $D$ (this implies in particular $(P\cap D)=1$). The Legendre symbol \cite{serre,hardy} is then defined as:
\begin{eqnarray}
\label{legendre}
\left({D\over P}\right)&=&1\quad\mbox{if $P$ splits w.r. to $D$} \\
\left({D\over P}\right)&=&0\quad\mbox{if $P$ is a divisor of $D$}\nonumber\\
\left({D\over P}\right)&=&-1\quad\mbox{otherwise ($P$ is said to be inert w.r.to $D$)} \nonumber
\end{eqnarray}
 (In the following, we might forget the ``w.r.to $D$" when mentioning a splitting prime). 

The case where $P$ is inert will not be considered since it yields no eigenmodule. We will treat
the case of splitting primes in the main text, and defer two complementary cases ($P|D$ and $P=2$) to Appendix B.

 To go from prime to 
primary values $N=P^\beta$, we will use the following lemma \cite{serre}:
\begin{lem}
\label{theorem}
To any solution $k_P$ of an algebraic equation $f(k)\equiv 0$ modulo $P$ s.t. $f'(k_P)\not\equiv 0$ mod $P$, there corresponds for any $\beta>1$ a unique solution $k_N$ modulo $N=P^\beta$; moreover, $k_P\equiv k_N$ mod $P$.
\end{lem}
 Finally, we will combine these results to deal with composite
values (i.e. products of admissible primaries).

\subsection{$N=P^\beta$, a power of an odd splitting prime}
\label{split} 
We consider the case $N=P^\beta$, with $P$ an odd prime number, $\left({D\over P}\right)=1$, and $\beta\geq 1$. If the vector $\vec v=\left(\begin{array}{c}n\\-m\end{array}\right)$ generates a free submodule $\Lambda$ in ${\Bbb Z}_N^2$, the condition $(n\cap m \cap N)=1$ implies that $n$ or $m$ is coprime with $P$, and thus with $N$. In the case $(n\cap N)=1$, $n$ admits an inverse $n^{-1}$ modulo $N$, such that $n^{-1}\vec v$ is also a generator of $\Lambda$. We thus have the following:
\begin{lem}
\label{form}
 For a primary value $N=P^\beta$, any free principal submodule $\Lambda$ of ${\Bbb Z}_N^2$ can be generated by a vector of the form
\begin{equation}
\label{forms} 
\left(\begin{array}{c}1\\k\end{array}\right)\quad\mbox{ or }\quad \left(\begin{array}{c}k'\\1\end{array}\right).
\end{equation}
\end{lem}
Then $\Lambda=\{n\vec v,\ n\in {\Bbb Z}_N\}$ is an eigenmodule of $S$ (which we will note $\Lambda = (\vec v,\lambda)$, with $\lambda$ the corresponding eigenvalue) iff the coefficient $k$ (resp. $k'$) is a solution of
\begin{eqnarray}
\label{kk'}
bk^2+(a-d)k-c&=&0\quad\mbox{mod }N\\
\mbox{(resp.)}\ \ \ \ ck'^2+(d-a)k'-b&=&0\quad\mbox{mod }N.\nonumber
\end{eqnarray}
Generally, $bc\not\equiv 0$ mod $P$, so both $b$ and $c$ are invertible modulo $N$; we can then solve (\ref{kk'}) provided $D$ has a square-root in $\Bbb Z_N$, using the same formulae (\ref{lambda}) as when working on $\Bbb R$. Since $P$ splits, there exist two such roots modulo $N=P^\beta$, for any $\beta\geq 1$, which we note $\pm x_N$ (cf. lemma \ref{theorem}). Each of the above equations has then two solutions $k_\pm=b^{-1}\{(d-a)/2\pm x_N\},k'_\pm =c^{-1}\{(a-d)/2\pm x_N\}$, associated to the eigenvalues $\lambda_\pm =(a+d)/2\pm x_N$. These solutions are of course redundant, since $k_\pm \,k'_\pm\equiv 1$ mod $N$. By convention, we can choose to generate
\begin{equation}
\label{lambdak}
\begin{array}{cccc}
 \mbox{the eigenmodule associated with}&\lambda_+&\mbox{ by the eigenvector}&\left(\begin{array}{c}1\\k_+\end{array}\right)\\
"&\lambda_-&"&\left(\begin{array}{c}k'_-\\1\end{array}\right)\end{array}.
\end{equation}

In the `caustic case' $bc\equiv 0$ mod $P$, a square-root is easily extracted modulo $P$,
\begin{eqnarray}
D\equiv \left(a+d\over 2\right)^2-1&\equiv& \left(a-d\over 2\right)^2\quad\mbox{mod }P\\
\Longrightarrow \mbox{we can take}\quad x_P&\equiv&{a-d\over 2}\quad\mbox{mod }P\nonumber
\end{eqnarray}
$(P\cap D)=1$ implies $(P\cap x_P)=1$. Therefore, each of the equations (\ref{kk'}) admits at least one solution modulo $P$, yielding two eigenspaces $\Lambda_{P,\pm}$, which can still be generated as in (\ref{lambdak}): the eigenvalue $\lambda_+=a$ (resp. $\lambda_-=d$) is associated with $k_+=c/(a-d)$ (resp. $k'_-=b/(d-a)$).  The extension from the case $\beta=1$ to any $\beta>1$ is done through the lemma \ref{theorem}.

Although it is very general, the choice (\ref{lambdak}) for generators is not compulsory; for instance, if $b$ is coprime to $N$, it is possible, and sometimes more convenient, to generate both eigenspaces by a vector of type $\left(\begin{array}{c}1\\k\end{array}\right)$.

\medskip
In Appendix B, we similarly classify the eigenmodules when $N=P^\beta$, first for $P$ an odd prime divisor of $D$ ($P|D\Leftrightarrow \left({D\over P}\right)=0$), then in the special case $P=2$ \cite{serre}. The analysis is slightly more involved than in the present section, but it only concerns a finite number of primes, which makes it much less `generic' in the large-$N$ limit. 

From the knowledge of free eigenmodules for primary values $N$, one easily deals with composite values.

\subsection{$N$ composite}
We treat the case $N=\prod_{i=1}^nN_i=\prod_{i=1}^nP_i^{\beta_i}$, with $P_i$ prime. If there exists a free eigenmodule $\Lambda_N=(\vec {v}_N,\lambda_N)$ modulo $N$, then the submodules $\Lambda_i=(\vec {v}_i,\lambda_i)$ modulo $N_i$, given by 
\begin{eqnarray}
\label{chinese}
 \lambda_i&\equiv&\lambda_N\quad\mbox{mod }N_i,\quad\forall i=1,\cdots, n\\
\vec{v}_i&\equiv&\vec{v}_N\quad "\qquad"\qquad"\qquad"\nonumber
\end{eqnarray}
are free and invariant, hence every factor $N_i$ is admissible.
The Chinese Remainder Theorem \cite{hardy,serre} yields the inverse property: 
given free eigenmodules $\Lambda_i=(\vec {v}_i,\lambda_i)$ modulo $N_i$, the equations (\ref{chinese}) have a unique solution $(\vec {v}_N,\lambda_N)$ modulo $N$, corresponding to a free eigenmodule $\Lambda_N$, which is independent of the generators we choose for the $\Lambda_i$'s; indeed, $\Lambda_N$ (although not $\vec{v}_N$) is unchanged if we replace $\vec{v}_i$ by $u_i\vec{v}_i$, where $u_i$ is any invertible element of $\Bbb Z_{N_i}$. 

 If both $(b\cap N)>1$ and $(c\cap N)>1$, the generator $\vec{v}_N=\left(\begin{array}{c}k_N'\\k_N\end{array}\right)$ cannot be written in one of the forms (\ref{forms}).
However, the coordinates $k_N',k_N$ can be chosen mutually prime: indeed, the free nature of the submodule $\Lambda_N$ is equivalent to $(k_N'\cap k_N\cap N)=1$. Therefore, $G=(k_N\cap k_N')$ is invertible in $\Bbb Z_N$, and $\Lambda_N$ admits as generator $\vec{v}_N/G$, the coordinates of which are coprime. 

On the contrary, if $(b\cap N)=1$ (resp. $(c\cap N)=1$), then, as in the previous section, $k'_N$ (resp. $k_N$) can be taken equal to $1$.  

\paragraph{Summary : Which $N$ are admissible ?}
   
Let us now summarize our classification (including the results of Appendix B). If we decompose the discriminant of $S$ as in \cite{percival},
$D=t^2\,D_{\rm sf}$, where the second factor is square-free, the following results hold:

-for any prime $P$ s.t. $\left({D_{\rm sf}\over P}\right)=1$, $P^\beta$ is admissible for any $\beta$ (in the case $P=2$, the splitting condition means $D_{\rm sf}\equiv 1$ mod $8$). We notice that $\left({D_{\rm sf}\over P}\right)=1$ is implied by $\left({D\over P}\right)=1$, but the converse is false. 

-for other primes, $P^\beta$ is admissible if $\left({D\over P}\right)=0$ and $\beta$ is small enough; these cases only represent a finite set of primes (see Appendix B). 

-a composite number $N=\prod_{i=1}^nP_i^{\beta_i}$ is admissible iff each $P_i^{\beta_i}$ is.

The number $c_N$ of invariant strings for an admissible primary $N=P^\beta$ depends upon $(t\cap P)$. In the simplest case, $\left({D\over P}\right)=1$,
$c_N=2$ for any $\beta>0$. For a composite $N=\prod_{i=1}^nN_i$, we have $c_N=\prod_{i=1}^n c_{N_i}$.

\medskip
How frequent are the admissible numbers among all integers, in the semi-classical limit ? According to Hardy and Wright \cite{hardy}, a large number $N$ is typically the product of few large primes (on the average, $N$ contains $\omega(N)\sim \log\log N$ prime factors). To be admissible, such a large $N$ has to factorize into exclusively splitting primes, which represent statistically half of all primes. Therefore, admissible numbers become scarce in this limit. 

\medskip
To illustrate our calculations, we represent some linear invariant strings for the transformation $S_{\rm fig}=\left(\begin{array}{cc}2&1\\3&2\end{array}\right)$, which is the simplest hyperbolic symplectic matrix in $\Gamma_{1,2}$ (cf \cite{keat1}). Equations (\ref{kk'}) then reduce to $k^2=3$ or $3k'^2=1$ modulo $N$. Since $b=1$, all eigenvectors will be written in the form
${\vec v} =\left(\begin{array}{c}1\\k\end{array}\right)$. For the discriminant $D=3$, the lowest splitting prime is $11$, but $2$ and $3$ are admissible, although their powers are not. In figure 2, we show three invariant sublattices $\Lambda_N/N$ on ${\Bbb T}^2$, respectively for $N=11,\ 33,\ 121$. These 
sublattices look well-distributed over $\Bbb T^2$ as $N$ increases. This remark leads us to the next section, where we investigate some semi-classical properties of invariant strings.

\section{Asymptotic properties of invariant strings}
\label{asymp} 
 We are interested in the way an $S$-invariant sublattice $\Lambda_N/N$ equidistributes over $\Bbb T^2$, for a large admissible $N$.
To start with a simple case, let us assume that a generator $\vec v$ of $\Lambda_N$ can be written $\vec v= \left(\begin{array}{c}1\\k\end{array}\right)$, where $k$ is a solution of (\ref{kk'}) (this is possible, for instance, if $(b\cap N)=1)$.
We consider a smooth observable $f({\vec x})$ on $\Bbb T^2$, and estimate its average over this sublattice. We are thus led to the sum
\begin{equation}
<f>_{\Lambda_N}={1\over N}\sum_{n=1}^N f(n\vec v/N)={1\over N}\sum_{n=1}^N\sum_{{\vec p}\in \Bbb Z^2}f_{\vec p}\;\e^{2i\pi{{\vec p}.n{\vec v}\over N}}
\end{equation}
by Fourier transforming the biperiodic observable. Permuting the sums, we see that
\begin{equation}
{1\over N}\sum_{n=1}^N \e^{2\i\pi{{\vec p}.n{\vec v}\over N}}=
\left\{\begin{array}{cc} 1& \mbox{if }{\vec p}.{\vec v}\equiv 0\bmod N\\0&\mbox{otherwise.}\end{array}\right.
\end{equation}
Combining the above condition of `constructive interference' with eq.(\ref{kk'}), we obtain 
\begin{equation}
\label{equaQ}
bp_1^2-cp_2^2+(d-a)p_1p_2\stackrel{\rm def}{=}Q(\vec p)\equiv 0\bmod N.
\end{equation}
For $\vec{p}\neq \vec{0}$, this equation cannot hold in $\Bbb Z$ (equivalently, the isotropic set $Q^{-1}(\{0\})$ only intersects $\Bbb Z^2$ at the origin); indeed, multiplying $Q(\vec p)$ by $b$ leads to
\begin{equation}
bQ(\vec p)=(p_1b+p_2{d-a\over 2})^2-p_2^2 D
\end{equation}
and we know that $D$ is not a perfect square in $\Bbb Q$ (due to the hyperbolicity of $S$). Therefore the equation (\ref{equaQ}) implies $\vec p=\vec 0$ or $|Q(\vec p)|\geq N$. If the largest eigenvalue of the quadratic form $Q$ is noted $q_{S}=\max |(b-c/2)^2\pm\sqrt{(b-c/2)^2+D}\,|$, the second condition cannot be satisfied when $\vec p$ is inside the disk of radius $\sqrt{N /q_{S}}$. We then obtain the estimate
\begin{equation}
\label{est1}
|\!<f>_{\Lambda_N}-\,f_{\vec 0} \,|\leq \sum_{|\vec p|\geq \sqrt{N/q_{S}}}|\,f_{\vec p}\,|.
\end{equation}
Equation (\ref{est1}) is meaningful if the  Fourier transform of $f$ is in $l_1 (\Bbb Z^2)$. However, since any continuous function $f\in{\cal C}(\Bbb T^2)$ can be {\it uniformly} approximated (i.e., for the $\|\,.\,\|_\infty$ norm) by trigonometric polynomials (see \cite{rudin}, p.91), we obtain the following result:
\begin{prop}

\label{wstar}
For any infinite sequence of S-invariant sublattices $\{\Lambda_N,\;N\,admissible\}$,

$\forall   f\in{\cal C}(\Bbb T^2),\quad \lim_{N\to\infty} <f>_{\Lambda_N}=f_{\vec 0}$.
\end{prop}

The rate of convergence of this limit depends on the smoothness properties of $f$. Indeed, if $f\in {\cal C}^\alpha(\Bbb T^2)$ (i.e. $f$ is $\alpha$ times differentiable), its Fourier coefficients are bounded as:
\begin{equation}
|\,f_{\vec p}\,|\leq {C_\alpha\over |{\vec p}\,|^\alpha}\quad\mbox{for }|{\vec p}|\geq 1
\end{equation}
where $C_\alpha$ depends on $f$. If $\alpha\geq 3$, we obtain
\begin{equation}
\label{est2}
f\in {\cal C}^\alpha (\Bbb T^2)\Longrightarrow |\!<f>_{\Lambda_N}-\,f_{\vec 0} \,|\leq {2\pi C_{\alpha}\over 2-\alpha}\left({q_{S}\over N}\right)^{\alpha-2\over 2}.
\end{equation}

\medskip
When the generator $\vec v$ cannot be written in the form (\ref{forms}), the same type of estimates still hold. We outline the few changes needed for this case. By construction (see last section), $N$ factorizes into two coprime numbers $N=M\,M'$, such that the eigenmodule $\Lambda_N$ admits a generator $\vec v$ verifying
\begin{equation}
\vec v \equiv \left(\begin{array}{c}1\\k\end{array}\right)\bmod M,\quad\mbox{ and}\quad\vec v \equiv \left(\begin{array}{c}k'\\1\end{array}\right)\bmod M'.
\end{equation}
When estimating $<f>_{\Lambda_N}$, the orthogonality condition ${\vec p}.{\vec v}\equiv 0$ modulo $N$ leads to:
\begin{equation}
 \left\{\begin{array}{c}Q(\vec p)\equiv 0\bmod M\\ Q(\vec p)\equiv 0\bmod M'\end{array}\right.\\
\Longleftrightarrow Q(\vec p)\equiv 0\bmod N.
\end{equation}
The property $Q(\vec p)=0\Longrightarrow \vec p=\vec 0$ still holds, so we obtain the same estimates as in eq.(\ref{est1},\ref{est2}). 

\bigskip  
Taking the average of $f$ over the lattices $\Lambda_N/N$ actually amounts to considering a set of atomic probability measures $\mu_N$ on $\Bbb T^2$; in this point of view, Proposition \ref{wstar} means that such a sequence of measures converges, in the weak-$*$ sense, to the Liouville measure on the torus, as $N\to\infty$, $N$ admissible. We are interested in these atomic measures because they more or less correspond to the zero patterns of the quantum eigenstates we plan to build in the next two sections. Proposition \ref{wstar} shows that the zeros of such eigenstates are uniformly distributed over ${\Bbb T}^2$ in the semi-classical limit, which constitutes the main result of this article. 

\paragraph{Remark}
We notice that the above uniformity estimates were obtained quite simply, and for all admissible $N$. On the contrary, the authors of \cite{DEGI} had to restrict $N$ to splitting primes, and the uniformity property they obtained for atomic measures used non-trivial estimates of Kloosterman sums. The reason for this discrepancy is that these authors were considering families of atomic measures supported by {\it individual periodic orbits} modulo $N$, whereas ours are supported by {\it whole eigenmodules}, which are unions of `ideal' periodic orbits \cite{percival}.

\section{From classical invariant strings to quantum eigenstates}
\label{from}
Now that we have studied some asymptotic properties of invariant strings, we choose one of these strings for an admissible value $N$ and express the quantum eigenstate with corresponding (lattice-like) zero pattern as a pure theta state in some invariant space ${\cal H}_{N,\kappa}$. We then do the same for another type of invariant sublattices, namely square sublattices, which appear when $N$ is a perfect square. In the subsequent section, we will then build a larger family of eigenstates for the same $N$, allowing their zero patterns to be {\it crystalline} instead of lattice-like.

\subsection{Eigenzeros forming a lattice, $N$ admissible}
\label{lattice}

We start from an eigenmodule $\Lambda_{\rm lin}=(\vec v, \lambda)$, obtained in section (\ref{classical}). We can then search
a corresponding {\it affine} invariant lattice $\Lambda$, i.e. find a {\it real} 
vector $\vec v_o$ s.t. $\Lambda=\vec{v}_o+\Lambda_{\rm lin}$ is itself invariant through $S$. We will not make
here an exhaustive list of such affine invariant strings, but rather restrict the entries of
$\vec v_o$ to be integer multiples of $N/2$. Each eigenstate we will exhibit will then belong to one of the four spaces ${\cal H}_{N,\kappa}$, with $\kappa=$0, $\hat{0}$, 1/2, \i/2. We divide the study according to the parity of $N$, for
reasons to become obvious.

Let us start by considering $N$ {\it odd} and admissible; we select a free eigenmodule $\Lambda_{\rm lin}$ with a generating vector $\vec{v}=\left(\begin{array}{c}k'\\k\end{array}\right)$, where $k$ and $k'$ are coprime. Since $N$ is odd, we may choose the entries $k$, $k'$ to be of opposite
parity. Indeed, suppose they are both odd; the following transformations 
yield a new generator of $\Lambda_{\rm lin}$ in the desired form,
\begin{equation}
\label{parity}
\vec v=\left(\begin{array}{c}k'\\k\end{array}\right)\longrightarrow
\left(\begin{array}{c}k'+N\\k\end{array}\right)\longrightarrow
{\vec v}={1\over G}\left(\begin{array}{c}k'+N\\k\end{array}\right)\qquad\mbox{where }
G=(k\cap (k'+N)).
\end{equation} 
In that case, we can find two integers $u$, $v$, s.t. the matrix
\begin{equation}
\label{Sv}
S_{\vec v}=\left(\begin{array}{cc}u&k'\\v&k\end{array}\right)
\end{equation}
belongs to $\Gamma_{1,2}$ (see eq. \ref{subgroup}). This matrix maps the vertical string $\Lambda_\infty$ generated by $\vec{v}_\infty=\left(\begin{array}{c}0\\1\end{array}\right)$ modulo $N$, onto
$\Lambda_{\rm lin}$. Since $\Lambda_\infty$ is associated to
the {\it antiperiodic} basis state $|q_{(N-1)/2}=1/2\rangle_{N,\hat{0}}$, the 
eigenstate associated with $\Lambda_{\rm lin}$ is itself antiperiodic, and reads as
\begin{equation}
|\psi\rangle_{N,\hat{0}}=U_{S_{\vec v}}\,|1/2\rangle_{N,\hat{0}}\ .
\end{equation}
The zero pattern of this state consists in the lattice $\overline{\Lambda}_{\rm lin}/N\sq2$, obtained from $\Lambda_{\rm lin}$ through a dilation and reflection w.r.to the real axis (or complex conjugation).
Since $N$ is odd, the four strings 
\begin{equation}
\label{lines}
\Lambda_{\rm lin},\ \ \Lambda'=\Lambda_{\rm lin}+\left(\begin{array}{c}N/2\\N/2\end{array}\right),\ \ 
\Lambda_1=\Lambda_{\rm lin}+\left(\begin{array}{c}N/2\\0\end{array}\right),\ \
\Lambda_\i=\Lambda_{\rm lin}+\left(\begin{array}{c}0\\N/2\end{array}\right)
\end{equation}
are all different. Since $S\in\Gamma_{1,2}$, $\Lambda'$ is invariant, and 
yields a periodic eigenstate
\begin{equation}
 |\psi'\rangle_{N,0}=U_{S_{\vec v}}|q_o=0\rangle_{N,0}\ .
\end{equation}
 If moreover $S$ is congruent to the identity matrix modulo $2$ (equivalently, $S\in\Gamma_2$, the level-2 principal congruence subgroup \cite{mumford}) , then $\Lambda_1$ and $\Lambda_\i$ are invariant as well, and yield respectively the
eigenstates
\begin{eqnarray}
 |\psi_1\rangle_{N,1/2}&=&U_{S_{\vec v}}|q_o=0\rangle_{N,1/2},\\
 |\psi_\i\rangle_{N,\i/2}&=&U_{S_{\vec v}}|1/2\rangle_{N,\i/2}.\nonumber
\end{eqnarray}

\medskip
When $N$ is {\it even} and admissible, the coordinates $k'$, $k$ of the generator ${\vec v}$ of $\Lambda_{\rm lin}$ can still be chosen coprime, but we cannot modify their parity through the trick of eq.(\ref{parity}), so the matrix $S_{\vec v}$ of equation (\ref{Sv}) is not always in $\Gamma_{1,2}$. Besides, 
the four sublattices given in eq.(\ref{lines}) are not all different. For example,
if $k,\ k'$ are odd, then $\Lambda_{\rm lin}=\Lambda'$, $\Lambda_1=\Lambda_\i$ are invariant and the corresponding eigenstates are both antiperiodic,
\begin{eqnarray}
|\psi\rangle_{N,\hat{0}}&=&U_{S_{\vec v}}|1/2\rangle_{N,1/2},\\
|\psi_1\rangle_{N,\hat{0}}&=&U_{S_{\vec v}}|0\rangle_{N,1/2}.\nonumber
\end{eqnarray}
Similar phenomena occur for different parities of $k$, $k'$. For instance, if $k'$ is even, then $\Lambda_{\rm lin}=\Lambda_1$, $\Lambda'=\Lambda_\i$ are both invariant, and the two eigenstates are in ${\cal H}_{N,1/2}$.  

\medskip
To summarize, in the case where $N$ is odd admissible, we can build from any invariant linear string $\Lambda_{\rm lin}$ both a periodic and
an anti-periodic eigenstate (plus two other eigenstates iff $S\in\Gamma_2$); when $N$ is even, we can build from $\Lambda_{\rm lin}$ two eigenstates which belong to the same space ${\cal H}_{N,\kappa}$, with the Floquet parameter $\kappa$ 
depending on the invariant string $\Lambda_{\rm lin}$ we consider. In figure 3, we draw the Husimi functions of two lattice-like periodic eigenstates of $U_{S_{\rm fig}}$, for the two admissible values $N=33$ and $N=121$; these eigenstates are built using the invariant strings drawn on figure 2 (center and right).

The following section deals with a slightly different type of eigenstates in the particular case when $N$ is a perfect square. 
 
\subsection{Eigenzeros on a square lattice}
\label{still}
Irrespectively of the number theory leading to free eigenmodules, a simple integer lattice shows up when $N$ is a perfect square, say $N=M^2$: namely, the square lattice $\Lambda_{\rm sq}=M(\Bbb Z +\i\Bbb Z)$, invariant through any modular transformation $S$, and containing $N$ points modulo $N$ ($\Lambda_{\rm sq}$ is also a submodule of ${\Bbb Z}_N^2$, but of type $2$ and not free).
Since the unique state of ${\cal H}_{1,\hat{0}}$ has a simple zero at the origin, the function $z\longmapsto \langle Mz|1/2\rangle_{1,\hat{0}}$ has zeros on the square lattice of side $1/M\sq2$, i.e. on $ \Lambda_{\rm sq}/N\sq2$. A straightforward calculation shows that this function is the Bargmann representation of a state $\langle z|\psi_{\rm sq}\rangle_{N,\kappa}$, periodic if $N$ is even, anti-periodic if $N$ is odd. Writing down the theta series, we easily get the decomposition of $|\psi_{\rm sq}\rangle_{N,\kappa}$ in the position basis,
\begin{equation}
 |\psi_{\rm sq}\rangle_{N,\kappa} ={1\over \sqrt{M}}\sum_{\mu=0}^{M-1}(-1)^\mu\  |{\mu+1/2\over M}\rangle_{N,\kappa}
\end{equation}
with $\kappa=0$ (resp. $\kappa={\hat 0}$) for $N$ even (resp. odd).
These states are not pure theta states according to the definition we gave in section \ref{eigen}, since they are not built through the action of ${\rm Mp}(2,\Bbb Z)$ onto some basis state of ${\cal H}_{N,\kappa}$. However, they transform simply, as shown below. Taking any transformation $S\in\Gamma_{1,2}$, we label the operator $U_S$ according to the space it acts upon:
\begin{equation}
\label{inv}
\langle z|U_S^{{\cal H}_{N,\kappa}}|\psi_{\rm sq}\rangle_{N,\kappa} = 
\langle Mz|U_S^{{\cal H}_{1,\hat{0}}}|1/2\rangle_{1,\hat{0}} =
\e^{2\i\pi\sigma_{\hat 0}}\langle Mz|1/2\rangle_{1,\hat{0}}
\end{equation}
with $\kappa=0$ or $\kappa=\hat{0}$ according to the parity of $N$. The first equality is due to the linearity of the transformations in eq.(\ref{linear}), the second is obvious (${\cal H}_{1,\hat{0}}$ is a one-dimensional space on which $U_S$ is an endomorphism). The eigenangle $\sigma_{\hat 0}$ only depends on the transformation $S$. 

\medskip
For the same $N=M^2$, we can build a second eigenstate associated to the invariant lattice $\Lambda_{\rm sq}'=\Lambda_{\rm sq}+\left(\begin{array}{c}M/2\\M/2\end{array}\right)$. The quantum state can be built as above, it is periodic for all $M$ :
\begin{equation}
|\psi_{\rm sq}'\rangle_{N,0}  = 
{1\over \sqrt{M}}\sum_{\mu=0}^{M-1} \ |\mu/M\rangle_{N,0}\ .
\end{equation} 

\begin{prop}

For any $S\in\Gamma_{1,2}$, for any perfect square $N$, the states $|\psi_{\rm sq}\rangle_{N,\kappa},\;|\psi_{\rm sq}'\rangle_{N,0}$ are eigenstates of $U_S$.
\end{prop}
In figure 4, we draw the Husimi function of the periodic eigenstate $|\psi_{\rm sq}'\rangle_{121,0}$ for $N=121$.  

\section{Crystalline eigenfunctions}
\label{crystal}

In the last section, we built eigenstates for which the zero patterns were classically-invariant lattices $\Lambda$, which filled the unit torus uniformly in the semi-classical limit. We would like to find a larger class of eigenstates with the same property.

The authors of \cite{DEGI} derived semi-classical equidistribution properties concerning the Wigner distribution of {\it complete} orthonormal families of eigenstates, although they had to restrict $N$ to splitting prime values to obtain these results. Equivalently, they proved that for these values of $N$, Schnirelman's theorem is valid for {\it any} sequence of eigenstates; their proof relies on the bilinearity of the semi-classical measures (Wigner of Husimi) w.r.to the quantum states.  

On the opposite, we are not directly concerned with the properties of the Husimi densities, but rather with the atomic measures associated to their zero patterns. Since these patterns are {\it very non-linear} objects, we are not able to generalize the above uniformity properties to complete families of eigenstates. However, we can obtain a partial generalization from last section: when $N$ is {\it not} a prime, it is sometimes possible to build eigenstates with zero patterns forming {\it crystals} instead of lattices, whose periodicities correspond to {\it non-free} eigenmodules of ${\Bbb Z}^2_N$.

 These crystalline states can be built through different ways, depending on the value of $N$. 

\subsection{$N$ admissible}
\label{admissible}
To simplify the presentation, we will give details only in the case where $N$ is odd admissible, and moreover we first consider an invariant linear string $\Lambda=\Lambda_{\rm lin}=(\vec v,\lambda)$ such that $\vec v$ is of the form $\left(\begin{array}{c}1\\k\end{array}\right)$, with eigenvalue $\lambda=a+bk$. Therefore, the matrix $S_{\vec v}$ of eq.(\ref{Sv}) can be written 
\begin{equation}
\label{Sv2}
S_{\vec v}=\left(\begin{array}{cc}0&1\\-1&k\end{array}\right).
\end{equation}
 We note the vertical integer translations of $\Lambda$ (which are, in general, not invariant under $S$):
\begin{equation}
\Lambda_{l}=\Lambda+\left(\begin{array}{c}0\\l\end{array}\right)=S_{\vec v}\left[ \Lambda_\infty +\left(\begin{array}{c}-l\\0\end{array}\right)\right],\quad\mbox{for } l\in{\Bbb Z}_N.
\end{equation}
 By linearity, $S$ maps the string $\Lambda_{l}$ into a parallel string $\Lambda_{l'}$. More precisely, $\Lambda_{l'}=\Lambda_{\lambda '\,l}$, where $\lambda'=d-bk$ is the inverse of $\lambda$ modulo $N$. To these translated strings are associated the anti-periodic pure theta states:
\begin{equation}
|\psi_l\rangle_{N,\hat{0}}=U_{S_{\vec v}}\;|1/2-l/N\rangle_{N,\hat{0}}\ .
\end{equation}
These states are not eigenvectors of $U_S$, but they form a new orthonormal basis of ${\cal H}_{N,\hat{0}}$, on which $U_S$ acts quite simply:
\begin{equation}
\label{newb}
U_S\,|\psi_l\rangle_{N,\hat{0}}=\e^{2\i\pi\sigma(N,l)}|\psi_{\lambda '\,l}\rangle_{N,\hat{0}}\ .
\end{equation}
The phases $\sigma(N,l)$ can be derived by writing down explicitly the matrix elements of $U_{S}$ in this new basis. This leads to more or less complicated formulas, some of which are given in Appendix A (see also \cite{eckhar,DE,DEGI} for the case $\kappa=0$). 

To build a basis of eigenstates, we use the partition of ${\Bbb Z}_N$ into disjoint cosets (or orbits) ${\cal O}_j=\{\lambda'^s\,l_j, s=0,\cdots, p_j-1\}$: $l_j$ is an arbitrary point on the orbit ${\cal O}_j$, and $p_j$ the period of the orbit. Using the usual notation $\phi(n)$ for Euler's totient function \cite{hardy}, we have necessarily $p_j|\phi(N_{l_j})$ (we recall that $N_l=N/(N\cap l)$).

Starting from an element $l_j$ of the coset ${\cal O}_j$, we can build a family of $p_j$ {\it orthonormal} eigenstates of $U_S$. According to eq.(\ref{newb}), $|\psi_{l_j}\rangle_{N,\hat{0}}$ and its successive images under $U_S$ are all eigenstates of $U_S^{p_j}$, with the same eigenvalue $\exp(2\i\pi\sigma_j)$, where
\begin{equation}
\sigma_j=\sum_{s=0}^{p_j-1} \sigma(N,\lambda'^s\,l_j).
\end{equation}
We therefore obtain $p_j$ anti-periodic eigenstates $|\phi_{j,r}\rangle_{N,\hat{0}}$ of $U_S$, with eigenvalues $\e^{2\i\pi (r+\sigma_j)/ p_j}$, for $r=0,\cdots, p_j-1$:
\begin{eqnarray}
\label{eigenphi}
|\phi_{j,r}\rangle_{N,\hat{0}}&=&{1\over \sqrt{p_j}}\;\sum_{s=0}^{p_j-1}c_s^{(r)}\, |\psi_{\lambda'^s\,l_j }\rangle_{N,\hat{0}}\\
\mbox{with}\qquad\forall r=0,\cdots, p_j-1,&&\quad c_o^{(r)}\mbox{ arbitrary, and}\nonumber\\
\forall s=0,\cdots, p_j-2,\qquad c_{s+1}^{(r)}&=&\e^{-2\i\pi(r+\sigma_j)/ p_j}\,
e^{2\i\pi\sigma(N,\lambda'^s\,l_o)}\;c_s^{(r)}.\nonumber
\end{eqnarray}
To each coset ${\cal O}_j$ thus corresponds a set of $p_j$ orthonormal eigenstates, with the eigenvalues written above. Two eigenstates built from different cosets are orthogonal, since they involve disjoint sets of orthonormal states $|\psi_l\rangle_{N,\hat{0}}$. However, they can share the same eigenvalue \cite{DEGI,keat2}. By completing the above procedure for all cosets ${\cal O}_j$ (including the trivial one ${\cal O}_0=\{0\}$), we obtain an orthonormal basis of (possibly degenerate) eigenstates.
 
\medskip
In general, these eigenstates are not pure theta states, and their zero patterns are not known analytically. However, if $N$ is {\it not a prime} and $(l_j\cap N)>1$ (this gcd does not depend on the element $l_j$ we have selected in ${\cal O}_j$), then the Bargmann functions of the eigenstates $|\phi_{j,r}\rangle$, $r=0,\cdots,p_j-1$ inherit partial quasi-periodicity properties from their components $\langle z|\psi_l\rangle$. To prove this, we generalize the fine periodicity of basis states (eq. (\ref{fine})) to all pure theta states:
\begin{equation}
\label{finebis}
\langle z+{-b+\i d\over N\sq2}|U_S|q_j\rangle_{N,\kappa} = \langle z|U_S|q_j\rangle_{N,\kappa}\ \e^{-\pi(b+\i d)\sq2 z}\,\e^{\pi(b^2+d^2)\over 2N}\ \e^{2\i\pi q_j}.
\end{equation}   
If we apply this formula to the components of $|\phi_{j,r}\rangle$ and iterate it, we obtain
\begin{equation}
\langle z+\mu{-1+\i k\over N\sq2}|\psi_{\lambda'^s\,l_j}\rangle_{N,\hat{0}}=\langle z|\psi_{\lambda'^s\,l_j}\rangle_{N,\hat{0}}\ (-1)^{\mu}\ \e^{-\pi\mu(1+\i k)\sq2 z}\,\e^{\mu^2\pi(1+k^2)\over 2N}\,\e^{-{2\i\pi\mu\lambda'^s\,l_j\over N}}.
\end{equation}
The only $s$-dependent factor is the last one. If we choose $\mu=N_{l_j}$, this dependence is removed, so that the eigenstates $|\phi_{j,r}\rangle_{N,\hat{0}}$ themselves share the above periodicity property,
\begin{equation}
\label{periodicity}
\langle z+N_{l_j}{-1+\i k\over N\sq2}|\phi_{l_j,r}\rangle_{N,\hat{0}}=\langle z|\phi_{l_j,r}\rangle_{N,\hat{0}}\ (-1)^{N_{l_j}}\,
\e^{-\pi N_{l_j}(1+\i k)\sq2 z}\,\e^{\pi N_{l_j}^2(1+k^2)/ 2N}.
\end{equation}
As a consequence, the $N$ zeros of $\langle z|\phi_{j,r}\rangle_{N,\hat{0}}$ form a {\it crystal}, whose lattice is generated by the vector (in complex representation) $(-1+\i k)/(N\cap l_j)\sq2$, and contains $(N\cap l_j)$ points; each fundamental cell contains $N_{l_j}$ zeros, but we have no information about their locations within the cell. 

If we rename our basis of eigenstates as
\begin{equation}
\label{Phi}
|\Phi_{l_j \lambda'^r}\rangle_{N,\hat{0}}\stackrel{{\rm def}}{=}|\phi_{j,r}\rangle_{N,\hat{0}}\ ,
\end{equation}
we obtain a basis of orthonormal eigenstates $\{|\Phi_{l}\rangle_{N,\hat{0}},\ l=0,\cdots, N-1\}$ s.t. each zero pattern of $\langle z|\Phi_{l}\rangle_{N,\hat{0}}$ is a crystal generated by $(-1+\i k)/(N\cap l)\sq2$.

Note that if we perform the same construction of eigenstates using a different invariant string, we will a priori obtain a different orthonormal eigenbasis; this is not inconsistent, considering the frequent degeneracies of the spectrum \cite{keat1}. There can actually be redundancies between two eigenstate bases built from two different eigenmodules $\Lambda^{(1)}$, $\Lambda^{(2)}$. Indeed, if, for a certain divisor $g$ of $N$, we have ${\vec v}^{(1)}\equiv{\vec v}^{(2)}\bmod N/g$, then the crystals of the eigenstates $|\Phi^{(1)}_{l_1}\rangle$, $|\Phi^{(2)}_{l_2}\rangle$ with $g|l_1$, $g|l_2$, are both supported by the same lattice; if these eigenstates share the same eigenvalue, nothing prevents them from being rigorously identical.

\medskip
 To deal with an invariant sublattice with generator $\vec v=\left(\begin{array}{c}k'\\k\end{array}\right)$ (i.e. the case where both $(b\cap N)>1$ and $(c\cap N)>1)$, we cannot restrict ourselves to vertical translations of $\Lambda$, as was done so far. To build a complete eigenstate basis, we have to consider all possible translations of $\Lambda$ of the form ${\vec v}_o+\Lambda$, with ${\vec v}_o$ an integer vector. Thus, ${\vec v}_o$ should take values in a fundamental cell of the sublattice $\Lambda$, i.e. a representative of the quotient $(N\Bbb Z)^2/\Lambda$. Such a fundamental cell can be obtained by the same procedure as above, i.e. by $S_{\vec v}$-transforming (using eq. (\ref{Sv})) the vertical strings $\Lambda_\infty+\left(\begin{array}{c}l\\0\end{array}\right)$, for $l=0,\cdots,N-1$. The construction of eigenstates then proceeds as above, using the inverse eigenvalue $\lambda'=k(au+bv)-k'(cu+dv)$.    

\medskip
This construction can be generalized straightforwardly to obtain a basis of periodic ($\kappa =0$) eigenstates for the same odd $N$, by using integer translations of the string $\Lambda'$ (see section \ref{lattice}). One obtains a basis of orthonormal eigenstates $|\Phi'_l\rangle_{N,0}$ similar to the one obtained above (eq. (\ref{Phi})). 

As an illustration, figure 5 shows the Husimi functions of two crystal eigenstates of $U_{S_{\rm fig}}$ for $N=33=3\times 11$ (both $3$ and $11$ are admissible). On the left, the zero pattern is periodic w.r. to a lattice of order 11, whereas on the right, the periodicity lattice is of order 3.
 
To deal with an even admissible $N$, one just needs to take care of the Floquet parameter $\kappa$ the same way as in section \ref{lattice}.

\bigskip
Let us turn back to the orthonormal basis of antiperiodic eigenstates $\{|\Phi_l\rangle_{N,\hat{0}}, \; l\in {\Bbb Z}_N\}$. Noting $g=(N\cap l)$, the state  $|\Phi_l\rangle_{N,\hat{0}}$ has a zero pattern of period $(-1+\i k)/g\sq2$, which generates an invariant string of $g$ points on $T_{\Bbb C}$. Through a dilation of factor $g\sq2$, this linear string is mapped into a free principal eigenmodule of ${\Bbb Z}_g^2$. Averaging an observable $f$ over the crystal of zeros, we can therefore apply equidistribution estimates identical to eq.(\ref{est1},\ref{est2}), after replacing $N$ by $g$ in the formulas.
The zero patterns of $\langle z|\Phi_l\rangle_{N,\hat{0}}$ are thus well distributed over $T_{\Bbb C}$ when $g$ is large. Unfortunately, the eigenstates featuring such regular patterns are not very numerous: for any divisor $g$ of $N$, the number of integers $l\in{\Bbb Z}_N$ s.t. $(N\cap l)=g$ is given by $\phi(N/g)$, and 
Euler's totient function varies like $\phi(n)\sim 6n/\pi^2$ on average \cite{hardy}. As a matter of fact, the eigenstates $|\Phi_l\rangle_{N,\hat{0}}$ with no a priori crystal periodicity (i.e, $N$ and $l$ coprime) represent a non-negligible fraction $\phi(N)/N$ of the whole basis. 

\medskip
Finally, we notice that the quasi-periodicity of the Bargmann eigenfunction (\ref{periodicity}) is mapped to an {\it exact periodicity} of the corresponding Husimi density. Therefore, the uniformity estimates (\ref{est1},\ref{est2}) apply as well if one averages $f$ over the Husimi density itself. From the results above, we therefore deduce the following proposition: 
\begin{prop}
\label{conv}

Consider an infinite sequence ${\cal N}=\{N\in {\Bbb N},\,N\,odd\,admissible\}$ and a corresponding sequence of crystal eigenstates $\{|\Psi\rangle_N=|\Phi_l\rangle_{N,\hat{0}}\}_{N\in{\cal N}}$ such that $g=(N\cap l)\to\infty$. Then both the Husimi measures $\{H_{\Psi_N}\}_{N\in{\cal N}}$ and the atomic measures $\{\mu_N\}_{N\in{\cal N}}$ describing the zero patterns of $H_{\Psi_N}$ converge weak-$*$ to the Liouville measure on ${\Bbb T}^2$ as $N\to\infty,\,N\in{\cal N}$.
\end{prop}

\subsection{$N$ a perfect square}
We consider again the case $N=M^2$, and construct an orthonormal family of periodic ($\kappa=0$) eigenstates. In section \ref{still}, we built the eigenstate associated with the lattice $\Lambda_{\rm sq}'=M(\Bbb Z+i\Bbb Z)+M/2(1+\i)$,
\begin{equation}
\langle z|\psi_{\rm sq}'\rangle_{N,0} = \langle Mz|q_0=0\rangle_{1,0}\ .
\end{equation}
We generalize the above formula to include states with zeros on translates of $\Lambda_{\rm sq}'/N\sq2$. 
One easily checks that the state $|\psi\rangle_N$ defined by $\langle z|\psi\rangle_N=\langle Mz|q_0=\kappa_2\rangle_{1,\kappa}$ belongs to ${\cal H}_{N,0}$ iff $\kappa$ is of the form $\kappa=(n_1+\i n_2)/M$ with $n_1,n_2\in{\Bbb Z}_M$; its zero pattern is associated to $\Lambda_{\rm sq}'+\left(\begin{array}{c}n_2\\n_1\end{array}\right)$. This family of periodic states is globally invariant under $U_S$, according to a generalization of eq.(\ref{inv}),
\begin{equation}
\langle Mz|U_S^{{\cal H}_{1,\kappa}}|\kappa_2\rangle_{1,\kappa} =
\e^{2\i\pi\sigma_\kappa}\langle Mz|\kappa_2'\rangle_{1,\kappa'},\qquad 
\mbox{with }\kappa'\equiv S\,\kappa\ \bmod [1,\i ]. 
\end{equation}
Moreover, these states are mutually orthogonal (this can be seen by decomposing them on both position and momentum bases in ${\cal H}_{N,0}$), so they form an orthonormal basis of ${\cal H}_{N,0}$.
Starting from a certain $\kappa^{(0)}$ of the above type, member of the coset ${\cal O}_j=\{S^i\,\kappa^{(0)}=\kappa^{(i)},\ i=0,\cdots,p_j-1\}$, we can then build $p_j$ periodic orthonormal eigenstates $|\phi_{j,r}\rangle_{N,0}$ of $U_S$, involving linear combinations of $\langle Mz|\kappa_2^{(i)}\rangle_{1,\kappa^{(i)}}$. As in last section, we finally obtain an orthonormal basis of eigenstates.  

If the coordinates of $\kappa^{(0)}$ have the gcd $(n_1\cap n_2\cap M)=g$, then all the states $|\kappa_2^{(i)}\rangle_{1,\kappa^{(i)}}$ share the following quasi-periodicity relations:
\begin{eqnarray}
  \langle M(z+1/g\sq2)|\kappa_2^{(i)}\rangle_{1,\kappa^{(i)}}&=&
\e^{{\pi\over 2}(M/g)^2+\pi\sq2 z(M^2/g)}\;\langle Mz|\kappa_2^{(i)}\rangle_{1,\kappa^{(i)}}\\
\langle M(z+i/g\sq2)|\kappa_2^{(i)}\rangle_{1,\kappa^{(i)}}&=&
\e^{{\pi\over 2}(M/g)^2-\i\pi\sq2 z(M^2/g)}\;\langle Mz|\kappa_2^{(i)}\rangle_{1,\kappa^{(i)}}\ .\nonumber
\end{eqnarray}  
Since these relations are $(i)$-independent, they apply to each eigenstate $|\phi_{j,r}\rangle_{N,0}$ as well, the $N$ zeros of which belong to a crystal supported by the square sublattice in $T_{\Bbb C}$ of side $1/g\sq2$ (each fundamental cell of the crystal contains $(M/g)^2=N/g^2$ zeros).

In figure 6 (left) we plot the Husimi function of an eigenstate of $U_{S_{\rm fig}}$ for the square value $N=81$. The starting Floquet parameter is $\kappa^{(0)}=(3-3\i)/9$, so we obtain a square crystal of side length $1/3$.

In the case $N=M^2$ odd, a similar construction leads to anti-periodic eigenstates with the same type of zero patterns. 

As in the former subsection, the crystal of zeros will be well-distributed on the torus when
$g$ is large, which happens for relatively few eigenstates of the basis (namely, when both $n_1$ and $n_2$ are multiples of $g$).

\subsection{$N$ with an admissible divisor}
Since we are in search of families of eigenstates displaying semi-classical (i.e. $N\to\infty$) properties, the admissibility of $N$ seems a very restrictive condition. Indeed, this property implies that all prime divisors of $N$ are admissible themselves, which represents asymptotically half of all
primes. Therefore, the admissible numbers form
a set of vanishing density among all integers in the semi-classical limit. For this reason, we now use the ideas of the former sections to build crystalline eigenstates in the case
where $N$ is not admissible itself, but has an admissible divisor $M$, say $N=M\,m$ ($M$ is supposed maximal). We will build eigenstates at this value $N$, for which the zero pattern verifies the same periodicities as eigenstates built at the value $M$ (we restrict ourselves to odd $N$ and anti-periodic ($\kappa={\hat 0}$) states, which are the most amenable cases).

\medskip
The construction still uses classical results. We start from a free eigenmodule $\Lambda$ modulo $M$, generated by $\vec v=\left(\begin{array}{c}1\\k\end{array}\right)$ (we choose this form for $\vec v$ to simplify notations, so that the matrix $S_{\vec v}$ can be written as in  equation (\ref{Sv2})). Obviously, the lattice $m\Lambda$  is invariant modulo $N$ (it is a non-free principal eigenmodule of $S$ in ${\Bbb Z}^2_N$). We now consider a free module $\Lambda^o$ in ${\Bbb Z}^2_N$, which contains $m\Lambda$ as a submodule (for instance, we can take the string generated by the same $\vec v$ modulo $N$). $\Lambda^o$ is not invariant under
$S$, but its successive images $\Lambda^n=S^n\,\Lambda^o$ all contain $m\Lambda$. We note $p_{\rm rot}$ the least integer s.t. $S^{p_{\rm rot}}\Lambda^o=\Lambda^o$ ($p_{\rm rot}$ is typically of order $m$). This period corresponds to a cycle of `rotations' of $\Lambda^o$ under $S$, as opposed to the cycles of translations considered in the last sections. 

\medskip
Quantum-mechanically, the state
\begin{equation}
|\psi^o\rangle_{N,\hat{0}}=U_{S_{\vec v}}|1/2\rangle_{N,\hat{0}}
\end{equation}
and its successive images $U_S^n|\psi^o\rangle_{N,\hat{0}}\ $ are
pure theta states associated respectively to the lattices $\Lambda^n$. As a consequence, each of the states ${U_S}^n|\psi^o\rangle_{N,\hat{0}}$ has $M$ of its zeros on the string $\overline{\Lambda}/M\sq2$. Besides, ${U_S}^{p_{\rm rot}}|\psi^o\rangle_{N,\hat{0}}=\e^{2\i\pi\sigma_{\rm rot}}|\psi^o\rangle_{N,\hat{0}}$, the phase $\sigma_{\rm rot}$ depending on $S$ and $\Lambda^o$. We can then build a family of $p_{\rm rot}$ orthogonal eigenstates with eigenangles $\alpha_r={2\pi (r+\sigma_{\rm rot})\over p_{\rm rot}}$, for $r=0,\cdots,(p_{\rm rot}-1)$:
\begin{equation}
\label{proj}
 |\phi_r\rangle_{N,\hat{0}}={1\over \sqrt{p_{\rm rot}}}\;\sum_{n=0}^{p_{\rm rot}-1} \e^{-\i
 n\alpha_r}\;{U_S}^n|\psi^o\rangle_{N,\hat{0}}\ .
\end{equation}
This equation is similar to the formulas (\ref{newb},\ref{eigenphi}) we derived in the case of $N$ admissible. However, as opposed to section \ref{admissible}, these orthogonal eigenstates are {\it not normalized}, since the components $U_S^n|\psi^o\rangle_{N,\hat{0}}$ are not mutually orthogonal; some of the $|\phi_r\rangle_{N,\hat{0}}$ can even be null vectors, if $|\psi^o\rangle_{N,\hat{0}}$ is orthogonal to the corresponding eigenspaces (we expect this situation to be non-generic). 

The Bargmann functions of these $p_{\rm rot}$ eigenstates all vanish on $\overline{\Lambda}/M\sq2$, since all their components do. More precisely, these components (and thus all the eigenstates) possess the same quasi-periodicity w.r.to this sublattice:
\begin{equation}
\label{period0}
\forall n,\ \ \ \langle z+m{-1+\i k\over N\sq2}|{U_S}^n|\psi^o\rangle_{N,\hat{0}}=
\langle z|{U_S}^n|\psi^o\rangle_{N,\hat{0}}\;(-1)^m \e^{{\pi\over 2N}m^2(1+k^2)}\e^{-\pi m(1+ik)\sq2 z}.
\end{equation}

We have therefore built a family of orthogonal eigenstates whose zero patterns are crystals generated by ${-1+\i k\over M\sq2}$. For the inverse Planck's constant $M$, there was a unique state with such a property. 
In figure 6 (right), we plot the Husimi function of a periodic eigenstate of $U_{S_{\rm fig}}$ in the case $N=55$ ($11$ is admissible, $5$ is not); the invariant string $\Lambda$ modulo $11$ is shown in figure 2 (left). 

\bigskip
Further on, we can build eigenstates with coarser quasi-periodicities, by composing `rotations' and translations of $\Lambda^o$. Classically, the integer string $\Lambda^o+ \left(\begin{array}{c}0\\l_o\end{array}\right)=\Lambda^o+l_o\vec v_\infty$ modulo $N$ is mapped under successive actions of $S$ into sublattices of the form  $\Lambda^j+{\vec v}_j$, where ${\vec v}_j$ are integer vectors. We know that
$\Lambda^j$ will be a translate of $\Lambda^o$ iff $j$ is a multiple of
$p_{\rm rot}$. 

We can then use the results of section (\ref{admissible}) to describe
the cycle of translates of $\Lambda^o$ under the action of $S_{\rm rot}\stackrel{\rm def}{=}S^{p_{\rm rot}}$; this matrix maps 
$\Lambda^o+ l_o\vec v_\infty$ into $\Lambda^o+ \lambda'_{\rm rot}\,l_o\vec v_\infty$, where $\lambda'_{\rm rot}$ is an eigenvalue of $S_{\rm rot}$ modulo $N$. Without deriving explicitly the matrix $S_{\rm rot}$, we know that $\lambda'_{\rm rot}\equiv
\lambda'^{p_{\rm rot}}\bmod M$, where $\lambda'=d-bk$ describes the translates of $\Lambda$ modulo $M$ (cf. eq. (\ref{newb})). We can therefore decompose ${\Bbb Z}_N$ into cosets ${\cal O}_j=\{\lambda'^{s}_{\rm rot}l_j, s=0,\ldots,p_j-1\}$ (including the trivial coset ${\cal O}_0=\{0\}$). For each coset, the total period of the cycle $\{S^n\,(\Lambda^o+l_j\vec v_\infty)\}$ is thus $p_{\rm rot}\,p_j$. 

Considering the coset ${\cal O}_j$, the pure theta state we start with is $|\psi^o_{l_j}\rangle_{N,\hat{0}}=U_{S_{\vec v}}|1/2-l_j\rangle_{N,\hat{0}}$. Its images through $U_S$ are associated to the lattices $\Lambda^j+{\vec v}_j$, so they have a priori no common zero with 
$|\psi^o_{l_j}\rangle_{N,\hat{0}}$. However, as in section (\ref{admissible}), these states verify some non-trivial quasi-periodicity properties, cf. eq. (\ref{finebis}),
\begin{equation}
\label{fineter}
\langle z+m{-1+k\over N\sq2}|{U_S}^n|\psi^o_{l_j}\rangle_{N,\hat{0}}=
\langle z|{U_S}^n|\psi^o_{l_j}\rangle_{N,\hat{0}}\;(-1)^m \e^{{\pi\over 2N}m^2(1+k^2)}\e^{-\pi m(1+ik)\sq2 z}\e^{-2\i\pi {\lambda'^n\,l_j\over M}}, 
\end{equation}
and the only $n$-dependence appears in the last factor. From this sequence, $p_{\rm rot}\,p_j$ orthogonal eigenstates can be built using formula (\ref{proj}), with initial state $|\psi^o_{l_j}\rangle_{N,\hat{0}}$ and period $p_{\rm rot}\,p_j$ (the phase $\sigma$ now also depends on the coset ${\cal O}_j$). A priori, some of these states could be null, for the same non-orthogonality reasons as above. However, if such an eigenstate is not null and $g=(l_j\cap M)>1$, its zero pattern is a crystal generated by $(-1+\i k)/g\sq2$. 

For each coset ${\cal O}_j$, we have obtained a family of $p_j \,p_{\rm rot}$ orthogonal eigenstates (some of which can be null), with zeros on a crystal generated by $(-1+\i k)/(l_j\cap M)\sq2$. These crystals are based on the same lattices which appeared for the value $M$, but now each fundamental cell contains  $m(l_j\cap M)$ zeros. Unfortunately, the structure of the whole family of eigenstates is less clear than in section \ref{admissible}: on the one hand, we do not know the norms of the eigenstates, even their very existence; on the other hand, two degenerate eigenstates built from different cosets are a priori not mutually orthogonal. Actually, this lack of information is not surprising, since we have built $p_{\rm rot}$ times too many eigenstates through our procedure, so that dependence relations between them must exist. However, as in the last section, equation (\ref{fineter}) ensures that Proposition \ref{conv} applies to sequences of non-null eigenstates $|\Phi_{l_j}\rangle_{N,\hat{0}}$ for which $g=(l_j\cap M)\to\infty$.

\section{Concluding remarks}

Both the linearity of the transformation $S$ and the classical invariance
of certain sublattices (strings) of $\Bbb T^2$ transpose to
quantum mechanics, when expressed in the Bargmann--Husimi representation. On the time-dependent side, the quantum map $U_S$ acts classically on string-like zero patterns (i.e., zeros of pure theta states); on the time-independent side, it admits families of eigenstates for which the zero patterns are crystals supported by classically invariant sublattices.  

The lattices supporting these crystals (thus the crystals themselves) equidistribute on the torus when the number of lattice sites increases to infinity, and the equidistribution of the corresponding Husimi densities then results as a byproduct. 

However, crystalline eigenstates do not emerge for all values of $N$, and when they do, they concern only isolated states, whereas eigenspaces are very often multi-dimensional, due to arithmetical degeneracies of $U_S$. Obviously, a linear combination of two eigenstates with
crystals supported by {\it different} sublattices will not be crystalline, nor will their zeros be a priori well-distributed on $T_{\Bbb C}$: the equidistribution of the zeros is {\it not} a linear property. Therefore, the
eigenstates for which we have a nice zero pattern (i.e., a well-distributed crystal) only represent a few elements of the whole set of eigenstates; they do not even form a basis (indeed, they usually span a rather small subspace). Moreover, such crystal features certainly disappear as soon as one considers a non-linear perturbation of $S$.

In reverse, for many other eigenstates of $U_S$ that we have computed numerically, zeros show no particular periodicity, yet remain more or less well distributed on the torus. Surprisingly, this property persists even when the Husimi function of the eigenstate has a `scar' on a classical
fixed point (see figure 7). Therefore, the equidistribution property seems to be even more general for the zeros than for the Husimi density itself, but a proof will probably involve a different approach from that to Schnirelman's theorem.

\bigskip
{\bf Acknowledgements}
I would like to express my gratitude to A. Voros for numerous suggestions and advice, as well as a careful reading of the manuscript; I am also indebted to S. de Bi\`evre, M. Degli Esposti, S. Graffi, P. Leboeuf and S. Zelditch for fruitful discussions.

\vfill\eject

\appendix

\section*{Appendices}
\section{Derivation of the matrices $U_S$}
  
In this appendix, we derive the entries of the $N\times N$ matrix which represents the operator $U_S$ mapping the basis $\{|q_j\rangle_{N,\kappa},\ j=0,\cdots,N-1\}$ to the basis $\{|q_{j'}\rangle_{N,\kappa'},\ j'=0,\cdots,N-1\}$ (see eq. (\ref{qj})). The Floquet parameters $\kappa,\ \kappa'\in[0,1)^2$ are related by
\begin{equation}
\label{wind}
\left(\begin{array}{cc}a&b\\c&d\end{array}\right)\left(\begin{array}{c}\kappa_2\\ \kappa_1\end{array}\right)=\left(\begin{array}{c}\kappa_2'\\ \kappa_1'\end{array}\right)+ \left(\begin{array}{c}n_o\\m_o\end{array}\right),
\end{equation}
where the integers $n_o,m_o$ are called {\it winding numbers} \cite{keat2}. We will only consider hyperbolic matrices $S$, so the entry $b$ cannot vanish.
 The formulas obtained are generally used when $U_S$ is an endomorphism, i.e. $\kappa=\kappa'$, but we do not need this property in our calculations. For the periodic case ($\kappa=\kappa'=0$), the matrix elements were derived by
Berry and Hannay \cite{hannay:b} for any $N$. We recall how to obtain these elements for general Floquet parameters, and write explicit formulas in two cases, first when $N$ is coprime to $b$ (cf. our example $S_{\rm fig}$), then when $N$ divides $b$ (which is useful for section \ref{admissible}).

The Berry--Hannay approach consists in extending the action of $U_S$ from square-integrable functions to tempered quasi-periodic distributions ${\cal S}'(\Bbb R)$ (see also \cite{bouz}). Precisely, from the image of a unique Dirac peak,
\begin{equation}
U_S|q\rangle=\int dq'\ \langle q'|U_S|q\rangle\ |q'\rangle
\end{equation}
given by the kernel (\ref{kernel}), one derives the image of the quasi-periodic 
distribution $|q_j\rangle_{N,\kappa}$. From equation $(\ref{qj})$, we decompose the translation index as $m=b\nu + k$, with $\nu\in {\Bbb Z},\ k=0,\cdots,
b-1$, and obtain, using Poisson's summation formula,
\begin{equation}
U_S |q_j\rangle_{N,\kappa}={1\over \sqrt{\i Nb}}\sum_{n\in{\Bbb Z}}\ \sum_{k=0}^{b-1}\e^{{\i\pi \over b}(Nak^2-2nk)}\e^{{\i\pi N\over b}(dq'^2-2q_jq'+aq_j^2)}|q'=aq_j+b\kappa_1/N+n/N\rangle.
\end{equation}
In this equation, we already see that the image of $|q_j\rangle_{N,\kappa}$ is a sum of regularly spaced Dirac peaks. We now show that this image is itself quasi-periodic. For this matter, we compare the coefficients in front of $|q'_o+n/N\rangle$ and $|q'_o+n/N+1\rangle$, noting $q'_o=aq_j+b\kappa_1/N$. 

To avoid too cumbersome notations, we suppose from now on that $(b\cap N)=1$. The sum over $k$ then yields \cite{hannay:b}
\begin{equation}
\label{bcapN}
U_S |q_j\rangle_{N,\kappa}={C_N \over \sqrt{\i N}}\sum_{n\in{\Bbb Z}} \ 
\e^{-{\i\pi\over b}aN(aN\backslash b)^2n^2}\e^{{\i\pi N\over b}(dq'^2-2q_jq'+aq_j^2)}|q'=q'_o+n/N\rangle
\end{equation}
where $C_N$ is a number-theoretical phase (see eq.\ref{CN}), and $(aN\backslash b)$ is the integer inverse of $aN$ modulo $b$ (this inverse exists and is unique since $(aN\cap b)=1$). The exponents in front of $|q'_o+n/N+1\rangle$ and $|q'_o+n/N\rangle$
differ by
\begin{eqnarray}
-{\i\pi\over b}aN(aN\backslash b)^2(2nN+N^2) &+& {\i\pi N\over b}\left( 2d(q'_o+n/N)+d-2q_j\right)\\
&=&2\i\pi\kappa'_1+2\i\pi(cj+m_o)+{\i\pi\over b}(2n+N)(d-aN^2(aN\backslash b)^2).\nonumber
\end{eqnarray}
Finally, a simple argument shows that $\left(d-aN^2(aN\backslash b)^2\right)/b$ is an even integer. Therefore, we obtain a finite sum of quasi-periodic distributions,
\begin{equation}
\label{matrix}
U_S |q_j\rangle_{N,\kappa}={C_N \over \sqrt{\i N}}\sum_{n=0}^{N-1} \ 
\e^{-{\i\pi\over b}aN(aN\backslash b)^2n^2}\e^{{\i\pi N\over b}(dq'^2-2q_jq'+aq_j^2)}|q'=q'_o+n/N\rangle_{N,\kappa'}.
\end{equation}
To be complete, we give the value of the global factor $C_N$, 
\begin{eqnarray}
\label{CN}
\mbox{if $b$ is odd, } \ \ C_N&=&\left({aN\over b}\right)\e^{-\i\pi (b-1)/4}\\
\mbox{if $b$ is even, } \ \ C_N&=&\left({b\over aN}\right)\e^{\i\pi aN/4}\nonumber
\end{eqnarray}
where $\left({aN\over b}\right)$ is the Jacobi symbol, i.e. an extension of the Legendre symbol of eq.(\ref{legendre}) to non-prime $b$ (see \cite{hannay:b,lang} for details).
The unitarity of the matrix $U_S$ between the Hilbert spaces ${\cal H}_{N,\kappa}$ and ${\cal H}_{N,\kappa'}$ can be proven by decomposing $L_2(\Bbb R)$ into a direct integral of spaces ${\cal H}_{N,\kappa}$ \cite{bouz}: it then follows directly from the unitarity of $U_S$ acting on $L_2(\Bbb R)$.

Identifying each space ${\cal H}_{N,\kappa}$ with ${\Bbb C}^N$, we now express the matrix elements between the states $_{N,\kappa'}\langle q_{j'}|$ and $|q_j\rangle_{N,\kappa}$ for $j,j'=0,\cdots, N-1$. Due to the assumption $(b\cap N)=1$, $b$ admits a unique inverse $b^{-1}=(b\backslash N)$ modulo $N$. With this assumption, we can also choose the integer $b^{-1}$ s.t. $a\,b^{-1}$ and $d\,b^{-1}$ are even numbers. After a few calculations, eq.(\ref{matrix}) yields the following matrix elements,
\begin{eqnarray}
_{N,\kappa'}\langle q_{j'}|U_S|q_j\rangle_{N,\kappa}&=&
{\tilde C}_N \exp\left\{{2\i\pi\over N}\left[j'^2(d\,b^{-1}/2) -j'j\,b^{-1}+j^2(a\,b^{-1}/2)\right.\right. \\
&& \left.\left. +j'(\kappa'_1+m_o-d\,b^{-1}n_o)+j(b^{-1}n_o-\kappa_1)\right]\right\},\nonumber\\
\mbox{with the prefactor  }&& {\tilde C}_N={C_N\over\sqrt{\i N}}\exp\left\{{\i\pi\over Nb}(d\kappa_2'^2-2\kappa_2\kappa_2'+a\kappa_2^2)-{\i\pi \over b}aN(aN\backslash b)^2 n_o^2\right\}.\nonumber
\end{eqnarray}
In the periodic case $\kappa=\kappa'=0$, we recover an expression very 
similar in form to the one in the continuum, except for the prefactor (cf. \cite{hannay:b,DE}). Since we are mainly interested in $U_S$ being an endomorphism, we apply the above formula in the case $\kappa=\kappa'$, which is possible for a finite number of parameters $\kappa$, indexed by the winding numbers $n_o,m_o$ \cite{DE,bouz}:
\begin{equation}
\label{endom}
 \left(\begin{array}{c}\kappa_2 \\\kappa_1\end{array}\right)_{n_o,m_o}={1\over 2-a-d}\left(\begin{array}{cc}d-1&-b \\-c&a-1\end{array}\right)\left(\begin{array}{c}n_o \\m_o\end{array}\right).
\end{equation}
In the main text, we have only built eigenstates with periodicity angles $\kappa=0,1/2,\i/2,\,\hat{0}=(1+\i)/2$, but once given the matrix $S$, we can easily extend our results to all invariant parameters $\kappa_{n_o,m_o}$. 

\bigskip
 From eq.(\ref{bcapN}) on, the above calculations depended on the assumption $(b\cap N)=1$, which for instance is true for all values $N$ in the case $b=1$. If now $(b\cap N)>1$, the image of $|q_j\rangle_{N,\kappa}$ is still a superposition of quasi-periodic distributions in 
${\cal H}_{N,\kappa'}$, but now some elements $_{N,\kappa'}\langle q_j'|U_S|q_j\rangle_{N,\kappa}$ vanish. We are led to consider the case $N|b$ in section (\ref{admissible}), in the course of deriving the matrix elements $\langle\psi_{\lambda '\,l}|U_S|\psi_l\rangle_{N,\hat{0}}$. Indeed, 
group properties of $U_S$ lead to 
\begin{equation}
\langle\psi_{\lambda'\,l}|U_S|\psi_l\rangle_{N,\hat{0}}= \langle 1/2-\lambda '\,l|(\pm U_{S_{\vec v}^{-1}\,S\,S_{\vec v}})|1/2-l\rangle_{N,\hat{0}}
\end{equation}
(the $\pm$ sign does not depend on the bra-kets considered). For simplicity, we take $N$ odd, and suppose that $S_{\vec v}$ can be written $S_{\vec v}=
\left(\begin{array}{cc}0&1 \\-1&k\end{array}\right)$, with $k$ even. According to eq.(\ref{kk'}), the matrix ${\tilde S}=S_{\vec v}^{-1}\,S\,S_{\vec v}$ writes
\begin{equation}
{\tilde S}=\left(\begin{array}{cc}\lambda'=d-bk&\beta N \\-b&\lambda=a+bk\end{array}\right),
\end{equation}
where $\lambda$ is the classical eigenvalue associated with ${\vec v}=\left(\begin{array}{c}1\\k\end{array}\right)$ modulo $N$, $\lambda'$ its inverse, and $\beta=[bk^2+(a-d)k-c]/N$ is an integer coprime to $\lambda '$. By the same type of calculations as above, we obtain
\begin{equation}
_{N,\kappa'}\langle q_{j'}|U_{\tilde S}|q_j\rangle_{N,\kappa}=\sqrt{\beta\over \i}\e^{{\i\pi\over\beta}(\lambda q_{j'}^2-2q_{j'}q_j+\lambda' q_j^2)}\left\langle\exp\left\{{\i\pi\over\beta}[\lambda'm^2+2m(q_j \lambda'+\beta\kappa_1-q_{j'})]\right\}\right\rangle_m
\end{equation}
Here, $\left\langle.\right\rangle_m$ means the average over the integer $m$. This average is non-vanishing iff $q_j \lambda'+\beta\kappa_1-q_{j'}$ is integer, which corresponds, for each $j$, to a unique index $j'=j_S$ modulo $N$. Thus, the matrix $U_{\tilde S}$ acts, up to phase factors, as a permutation between both bases (cf eq.(\ref{newb})). Using the winding numbers 
\begin{equation}
{\tilde S}\,\kappa=\kappa'+ \left(\begin{array}{c}\tilde n\\\tilde m\end{array}\right),
\end{equation}
the state $|q_j\rangle_{N,\kappa}$ is mapped by $U_{\tilde S}$ into $\e^{\i\phi_j}|q_{j_S}\rangle_{N,\kappa'}$, where $j_S=\lambda' j+\tilde n$. The phase is given by
\begin{eqnarray}
_{N,\kappa'}\langle q_{j_S}|U_{\tilde S}|q_j\rangle_{N,\kappa}&=&
C_k\ \e^{\i\pi\beta\lambda\kappa_1^2}\ \ 
\e^{\i\pi Nb(\lambda'q_j^2-2q_j.q_{j_S})}\\  &=&C_k\ \e^{\i\pi\beta(\lambda\kappa_1^2-2b\kappa_1\kappa_2)}\e^{-\i\pi b\lambda'\kappa_2^2/ N}\ \e^{-{\i\pi b\over N}[\lambda'\,j^2+2j({\tilde n}+\kappa_2')]},\nonumber
\end{eqnarray}
where the constant $C_k=\left({\lambda'\over\beta}\right)\e^{-\i\pi\beta/4}$ if $b$ (and thus $\beta$) is odd, $C_k=\left({\beta\over\lambda'}\right)\e^{\i\pi(\lambda'+1)/4}$ if $b$ is even.
The formula simplifies greatly in the periodic case $(\kappa=\kappa'=0)$.

Since most of the eigenstates built in the text are {\it antiperiodic} $(\kappa=\kappa'=\hat{0})$, we now link the notations above to the ones in section (\ref{admissible}), i.e. we map the indices $j,j_S$ to the indices $l,l'$ of eq.(\ref{newb}), and adjust the sign of the image state:
\begin{eqnarray}
\mbox{\rm on the classical side}&&\mbox{\rm on the quantum side}\nonumber\\
j={N-1\over 2}-l&&
|q_j\rangle_{N,\hat{0}}\stackrel{\rm def}{=}|j/N+1/2N\rangle_{N,\hat{0}}=|1/2-l/N\rangle_{N,\hat{0}}\nonumber\\
j_S=\lambda'\,j+\tilde n,\ \ l'=\lambda'\,l&&
|q_{j_S}\rangle_{N,\hat{0}}=(-1)^{\lambda'+\beta-1\over 2}|1/2-l'\rangle_{N,\hat{0}}\ .
\end{eqnarray}

\section{Free eigenmodules, continued}
As a complement to section (\ref{split}), we classify the invariant strings modulo $N$, first when $N$ is a power of a prime divisor of $D$, then when $N$ is a power of $2$. We use the results of section (\ref{split}), in particular lemma \ref{form} still applies.

\subsection{$N=P^\beta$ with $P$ an odd prime, $P|D$}
Let $P\geq 3$ be a prime divisor of $D$, $\alpha>0$ be the largest integer such that $P^{\alpha}|D$. We seek to diagonalize $S$ modulo $N=P^\beta$, for $\beta \geq 1$.

 We first assume that $P$ does not divide $b$ and $c$ simultaneously, therefore {\it all} eigenmodules can be obtained by solving one of the eqs.(\ref{kk'}), so the problem reduces to extracting a square-root of $D$ modulo $N=P^{\beta}$, as in section (\ref{split}). 

For $\beta\leq\alpha$, the roots are the elements $x=X\,P^{[(\beta+1)/2]}$, where the brackets $[\ ]$ mean the integral part, $X$ taking any integer value. 

For $\beta>\alpha$, $D=P^{\alpha}\delta$ is a square in ${\Bbb Z}_{P^{\beta}}$ iff $\alpha$ is even {\it and} $\left({\delta\over P}\right)=1$. In those cases, its square-roots take the form $x=\pm y\,P^{\alpha/2}+X\,P^{\beta-{\alpha\over 2}}$, where $y$ verifies $y^2\equiv \delta\quad\mbox{mod }P^{\beta-\alpha}$, and
$X$ can be any integer.

In both cases, to each square-root $x$ modulo $P^\beta$ there corresponds one free eigenmodule $\Lambda$.

\medskip
In the case where $P$ divides $b$ and $c$ simultaneously, we decompose $S$ as 
\begin{equation}
S= \pm\left(\begin{array}{cc}1&0\\0&1\end{array}\right)+P^\gamma 
\left(\begin{array}{cc}e&f\\g&h\end{array}\right)=\pm I_2+P^\gamma M,
\end{equation}
with the condition $M\not\equiv 0$ mod $P$. We obtain the following constraints on $M$:
\begin{eqnarray}
\label{M}
{\rm det}S=1&\Longleftrightarrow& (e+h)=\mp \, P^\gamma(eh-gf)\\
 (e+h) \ \mbox{is even}&{\rm and}&D=P^{2\gamma}\left[\left({e+h\over 2}\right)^2-(eh-gf)\right]\nonumber
\end{eqnarray}
which implies $2\gamma\leq\alpha$. As long as $\beta\leq\gamma$, every vector of $\Bbb Z_N^2$ is invariant modulo $P^\beta$, so $\Lambda$ can be any free submodule of type one. For larger values of $\beta$, we need to diagonalize $M$ modulo $P^{\beta-\gamma}$, that is, to solve the remainders of eq. (\ref{kk'}),
\begin{eqnarray}
\label{kM}
fk^2+(e-h)k-g&\equiv& 0\quad\mbox{mod }P^{\beta-\gamma}\\
\mbox{or}\qquad gk'^2+(h-e)k'-f&\equiv& 0\quad\mbox{mod }P^{\beta-\gamma}.\nonumber
\end{eqnarray}
This amounts to extracting a square-root of the discriminant of $M$, $\delta'=D/P^{2\gamma}$ (this discriminant is related to the $\delta$ defined above by $\delta'=\delta.P^{\alpha-2\gamma}$). 

If $f$ or $g$ is coprime with $P$, the solutions of (\ref{kM}) will depend upon $\left({\delta'\over P}\right)$ : if $\left({\delta'\over P}\right)=1$, we apply the results of section (\ref{split}) and diagonalize $M$ for any $\beta$; if $P|\delta'$ (that is, $\alpha >2\gamma$), the first case considered in this appendix provides eigenvectors 

-at least up to $\beta=\alpha-\gamma$,

-for all $\beta$ iff $\alpha$ is even and $\left({\delta\over P}\right)=1$; 

If $\left({\delta'\over P}\right)=-1$, there is no eigenvector for $\beta >\gamma$. 

The case $f\equiv g\equiv 0$ mod $P$ is actually solved easily: since $\gamma$ is maximal, $e\equiv -h\not\equiv 0$ mod $P$, so the equations (\ref{kM}) have two solutions for any $\beta >\gamma$, using lemma \ref{theorem}.  

To conclude, we recall that any solution of (\ref{kM}) modulo $P^{\beta-\gamma}$ provides $P^\gamma$ distinct invariant strings modulo $P^\beta$.

\subsection{$N$ a power of $2$}
The prime number 2 has to be distinguished from the other primes, especially when one considers quadratic equations \cite{serre}, which play an important role in the diagonalization of $S$.
We separate the two possible forms of $S$ modulo $2$.

In the case $S\equiv \left(\begin{array}{cc}0&1\\1&0\end{array}\right)$, $S$ admits the unique eigenvector $\left(\begin{array}{c}1\\1\end{array}\right)$ mod $2$. Since $b$, $c$ are odd and $a-d$ even, eigenvectors are obtained by the extraction of a square-root of $D$ modulo $2^\beta$, for any $\beta>1$. 

If $(a+d)/2$ is even, then $D\equiv 3$ mod $4$ is not a square \cite{serre}, so there is no solution. 

If $(a+d)/2$ is odd, we write $D=2^\alpha \delta$, with $\delta$ odd, $\alpha\geq 3$. As long as $\beta\leq\alpha$, we have the trivial square-roots $x=X\,2^{[(\beta+1)/2]}$ for any integer $X$. When $\beta=\alpha+1$, there are solutions iff $\alpha$ is even, $x=2^{\alpha/2}+X\,2^{\alpha/2+1}$; for $\beta=\alpha +2$, the existence of solutions requires $\delta\equiv 1$ mod $4$, and for $\beta\geq\alpha+3$, the condition to extract a square-root is $\delta\equiv 1$ mod $8$. In the last two cases, all square roots are of the form $x=\pm 2^{\alpha/2}y + X\,2^{\beta-\alpha/2-1}$, where $y^2\equiv \delta$ mod $2^{\beta-\alpha}$.

\bigskip
The alternative case 
\begin{equation}
\label{M2}
S=\pm\left(\begin{array}{cc}1&0\\0&1\end{array}\right)+2^\gamma 
\left(\begin{array}{cc}e&f\\g&h\end{array}\right)=\pm I_2+2^\gamma M 
\end{equation}
leads to similar phenomena as for odd primes. 

For $\beta\leq\gamma$, all vectors  are trivially eigenvectors. 

For $\beta >\gamma$, if $f$ or $g$ is odd, eqs. (\ref{kM}) have solutions as long as the discriminant of $M$, $\delta'=\left({e+h\over 2}\right)^2-eh+gf$, admits square roots modulo $2^{\beta-\gamma}$; we are led to an analysis similar to the one above concerning $D=2^{2\gamma}\delta'$.

If both $f$ and $g$ are even, then $S\equiv (\pm 1+2^\gamma) I_2$ mod $2^{\gamma+1}$. This case implies $\alpha=2\gamma$, so $\delta=\delta'$. All vectors modulo $2^{\gamma+1}$ are eigenvectors. For $\beta\geq\gamma+2$, eq.(\ref{kM}) can be divided by a factor $2$, leading to:
\begin{eqnarray}
\label{kM2}
Fk^2+Ek-G\equiv 0\quad\mbox{mod }2^{\beta-\gamma-1} &&\mbox{or the analogous equation in }k',\nonumber\\
\mbox{where}\qquad F=f/2,\qquad E=(e-h)/2,&&G=g/2.
\end{eqnarray}
By definition, $M\neq 0$ mod $2$, so the equation (\ref{M}) implies that $e$, $h$, and $E$ are odd. Therefore, since $E$ is the derivative of the polynomial (\ref{kM2}) modulo $2$, lemma \ref{theorem} associates to each solution $k$ modulo $2$ a unique solution modulo $2^{\beta-1-\gamma}$, for any $\beta$. 
The analysis modulo $2$ is straightforward,
\begin{eqnarray}
(F,G)\equiv (0,0)&\Longrightarrow&k=0\quad {\rm or}\quad k'=0\\
(F,G)\equiv (1,0)&\Longrightarrow&k=0\quad {\rm or}\quad k=1\nonumber\\
(F,G)\equiv (0,1)&\Longrightarrow&k'=0\quad {\rm or}\quad k'=1\nonumber\\
(F,G)\equiv (1,1)&\Longrightarrow&\mbox{ no solution}\nonumber
\end{eqnarray}
Equivalently, in this last case ($f,\ g$ even) there are $2^{\gamma+2}$ solutions for any large $\beta$ if $\delta\equiv 1$ mod $8$ (the first three equations), and none if $\delta\equiv 5$ mod $8$ (the last equation), which are the only possible values of $\delta$ in this case.

\vfill\eject

\vfill\eject

\centerline{\bf Figure captions}
\bigskip
Fig. 1. Husimi function of the basis state $|q_2\rangle_{N=11,\kappa=0}$, plotted both in linear scale (gray levels; the high values are dark) and logarithmic (contour) scale. The linear scale shows a concentration of the density along the line $\{q=q_2=2/11\}$, whereas the logarithmic plot shows the vertical string of zeros along the line $\{q=q_2+1/2\}$. The function has the period $\i/11\sq2$ in the variable $z=(q-\i p)/\sq2$.

\medskip

Fig. 2. For the three admissible values of $N=11,\ N=33=11\times 3,\ N=121=11^2$, we plot free eigenmodules $\Lambda_N$ of $S_{\rm fig}=\left(\begin{array}{cc}2&1\\3&2\end{array}\right)$,
rescaled by $1/N$. The generating vectors are respectively
$\vec v_{11}=\left(\begin{array}{c}1\\6\end{array}\right),\ \vec v_{33}=\left(\begin{array}{c}1\\6\end{array}\right),\ \vec v_{121}=\left(\begin{array}{c}1\\94\end{array}\right)$. We notice that the string $\Lambda_{11}/11$ is a sublattice of $\Lambda_{33}/33$ (cf. eq. (\ref{chinese})), and of $\Lambda_{121}/121$ (cf. lemma \ref{theorem}). 

\medskip

Fig. 3. Husimi functions of two $\kappa=0$ pure theta eigenstates of $U_{S_{\rm fig}}$, for $N=33$ (left) and $N=121$ (right); we show the linear density scale using gray levels and indicate the position of the zeros by tiny circles. In both cases, the zero patterns are the rescaled affine strings $\overline{\Lambda'_N}/N\sq2$, where $\Lambda'_N$ are translates of the eigenmodules $\Lambda_{33}$, resp. $\Lambda_{121}$ of figure 2. For both eigenstates, the eigenangle is $-\pi/4$, of respective degeneracies $2$ and $3$.

\medskip

Fig. 4. Husimi function of the $\kappa=0$ eigenstate of $U_{S_{\rm fig}}$ associated with the invariant square sublattice $\Lambda'_{\rm sq}$, for $N=121$ (same representation as in fig. 3). The eigenangle is $-\pi/4$, hence this state is degenerate with the pure theta eigenstate shown in fig. 3 (right).

\medskip

Fig. 5. Husimi functions of two $\kappa=0$ cristalline eigenstates of $U_{S_{\rm fig}}$, for $N=33$. In both cases, we use translates of the eigenmodule $\Lambda_{33}$ of figure 2 (center). On the left, the coset ${\cal O}_1$ contains $l_1=11$, its period is $p_1=2$; the eigenangle is $2\pi/3-\pi/4$ (triply degenerate), and the crystal has the period $(-1+6\i)/11\sq2$. On the right, ${\cal O}_2$ contains $l_2=3$, its period is $p_2=10$; the eigenangle is $-2\pi/10-\pi/4$ (non-degenerate), and the crystal has the period $1/3\sq2$. The larger circles materialize double zeros.

\medskip

Fig. 6. Husimi functions of two $\kappa=0$ cristalline eigenstates of $U_{S_{\rm fig}}$. On the left, for the square value $N=81$, we use a cycle of translates of the lattice $\Lambda'_{\rm sq}$, starting from $\kappa^{(0)}=(3+3\i)/9$; this cycle is of period 6, the eigenangle is $2\pi/6-\pi/36$ (non-degenerate), and the zero pattern inherits a square periodicity of side $1/3\sq2$.
On the right, the value $N=55$ is not admissible, we use a cycle of rotations starting from a linear string $\Lambda^o$ containing the invariant sublattice $5\Lambda_{11}$ ($\Lambda_{11}$ is shown in figure 2 (left)). The cycle is of period $p_{\rm rot}=3$, yielding an eigenangle $2\pi/3-3\pi/4$ (four times degenerate), and the crystal has the period $(-1+6\i)/11\sq2$. Moreover, the function vanishes on the string $\overline{\Lambda'_{11}}/11\sq2$. 

\medskip

Fig. 7. Husimi function of two $\kappa=0$ eigenstates of $U_{S_{\rm fig}}$, for $N=33$ (left) and $N=59$ (right), both being admissible values. These state are built as described in subsection (\ref{admissible}), i.e. using translates of invariant strings. On the left, we use a translation cycle ${\cal O}_3$ of period $10$ containing $l_3=1$, the eigenangle is $2\pi/30-\pi/4$ (doubly degenerate); on the right, the cycle is of period $58$, and the eigenangle $\pi/4-2\pi/29$ is non-degenerate. In both cases, the state presents no particular periodicity, but rather a `scar' at a classical fixed point ($(1+\i)/2\sq2$ on the left, $0$ on the right). Nevertheless, the zero patterns seem well-distributed on the torus.  
\vfill\eject

\begin{figure}
\epsfxsize 17 truecm
\epsfbox{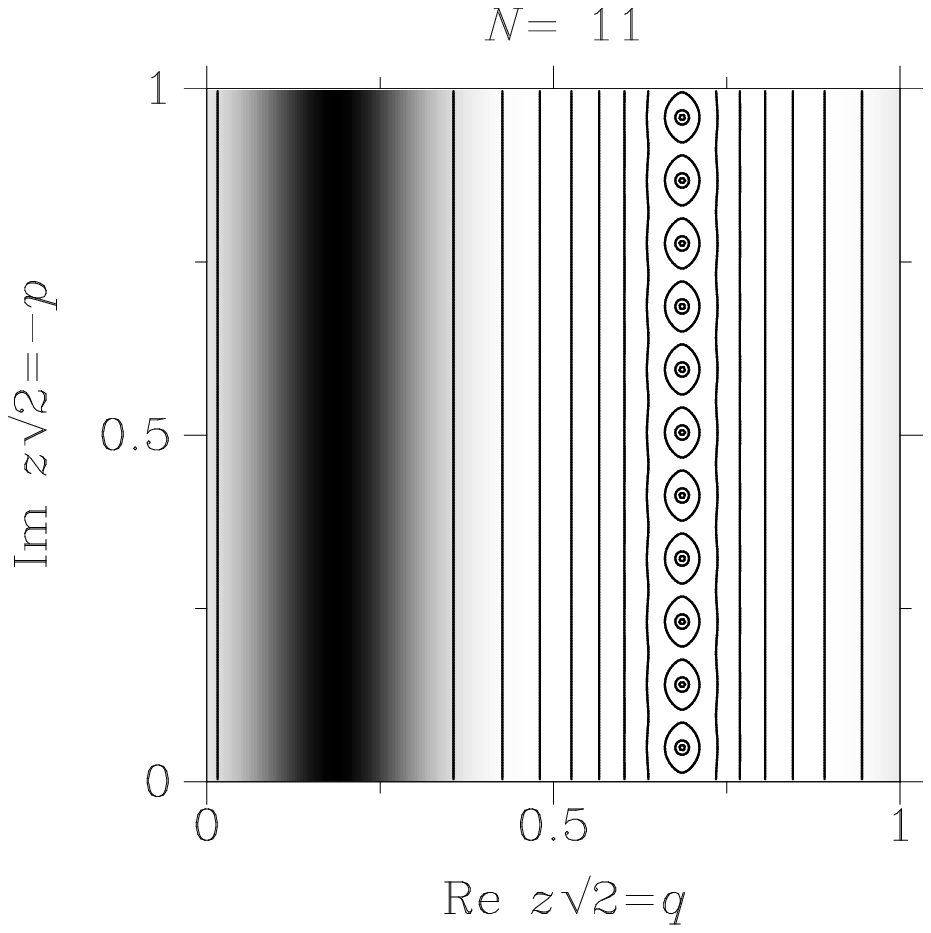}
\caption{}
\end{figure}
\vfill\eject
\begin{figure}
\epsfxsize 17 truecm
\epsfbox{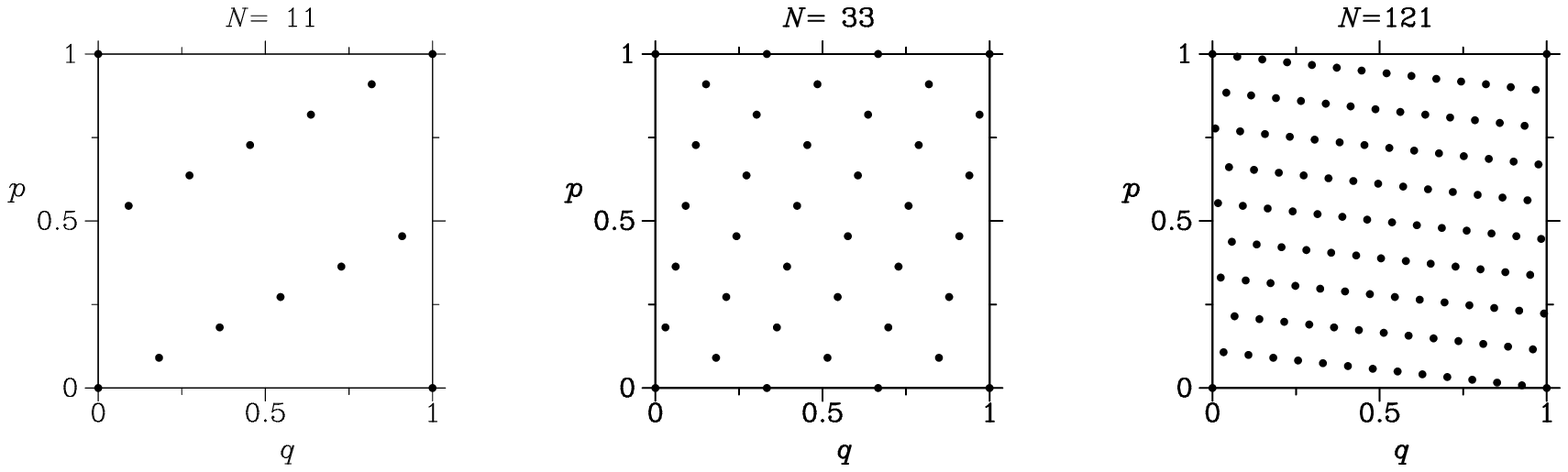}
\caption{}
\end{figure}
\vfill\eject
\begin{figure}
\epsfxsize 17 truecm
\epsfbox{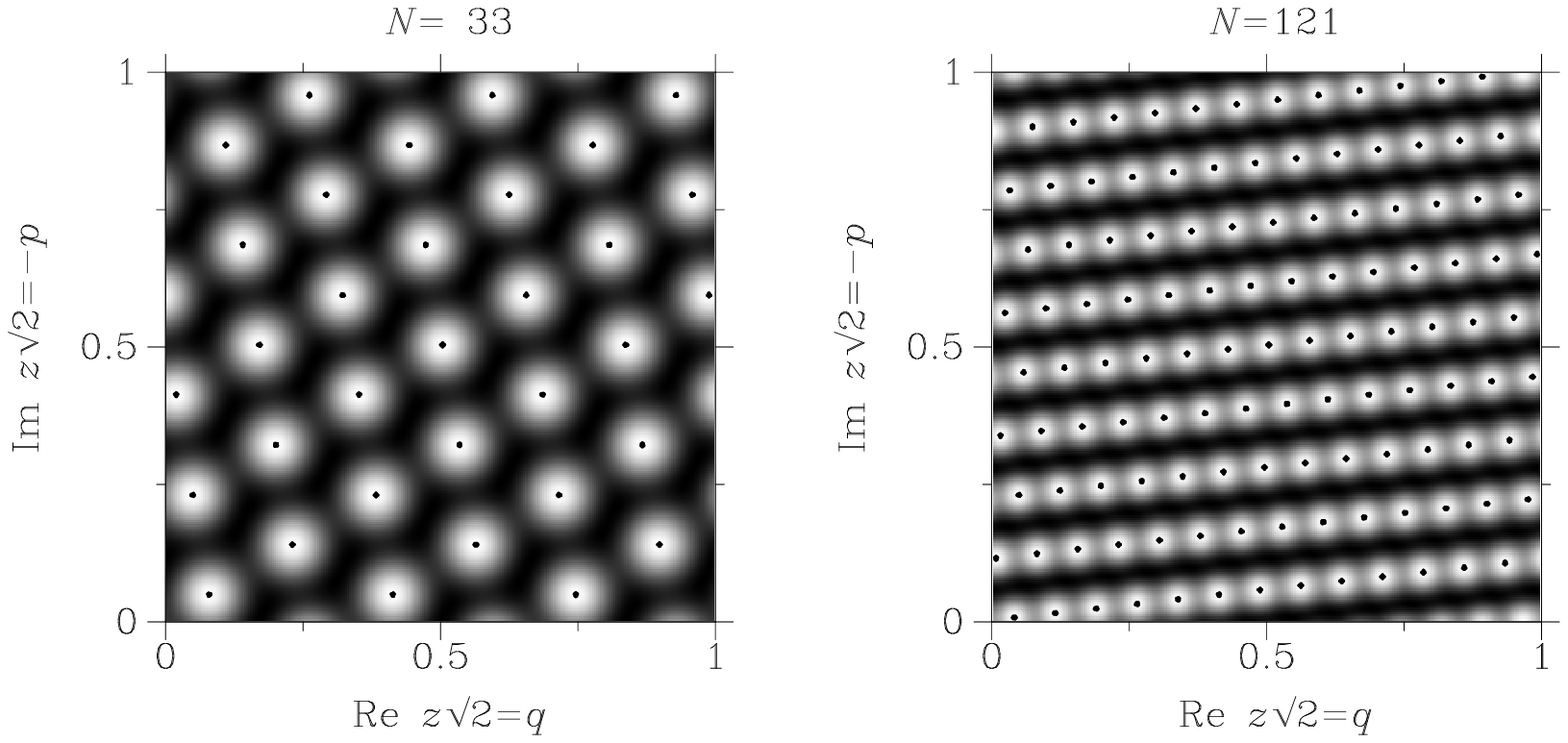}
\caption{}
\end{figure}
\vfill\eject
\begin{figure}
\epsfxsize 17 truecm
\epsfbox{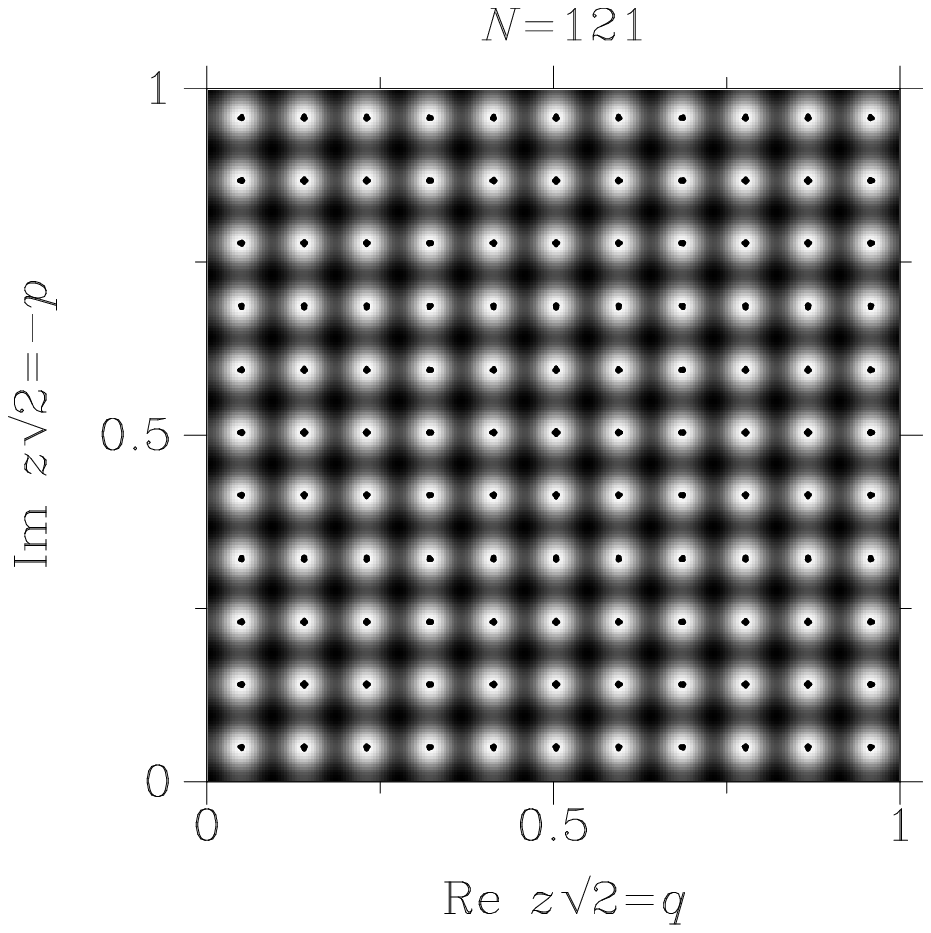}
\caption{}
\end{figure}
\vfill\eject
\begin{figure}
\epsfxsize 17 truecm
\epsfbox{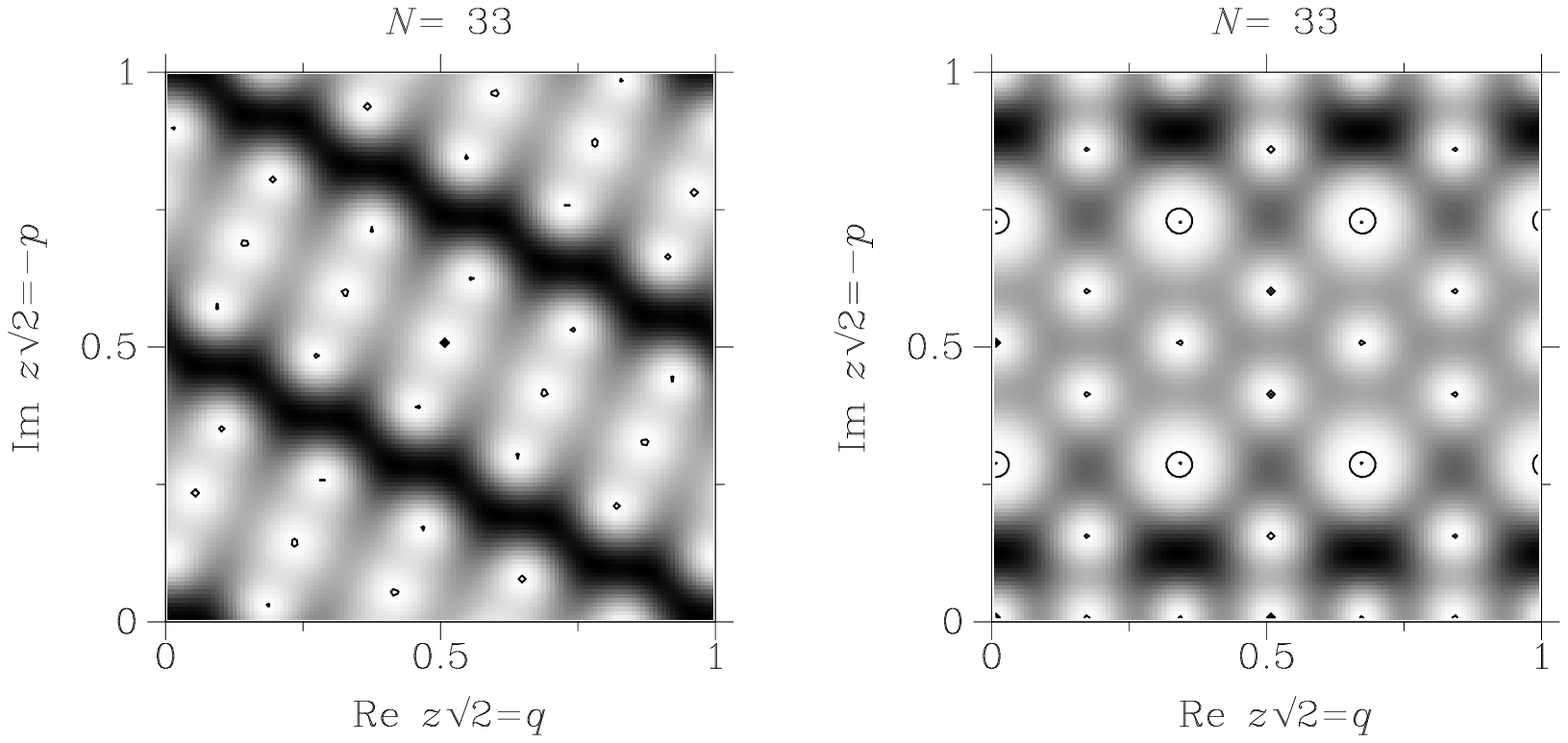}
\caption{}
\end{figure}
\vfill\eject
\begin{figure}
\epsfxsize 17 truecm
\epsfbox{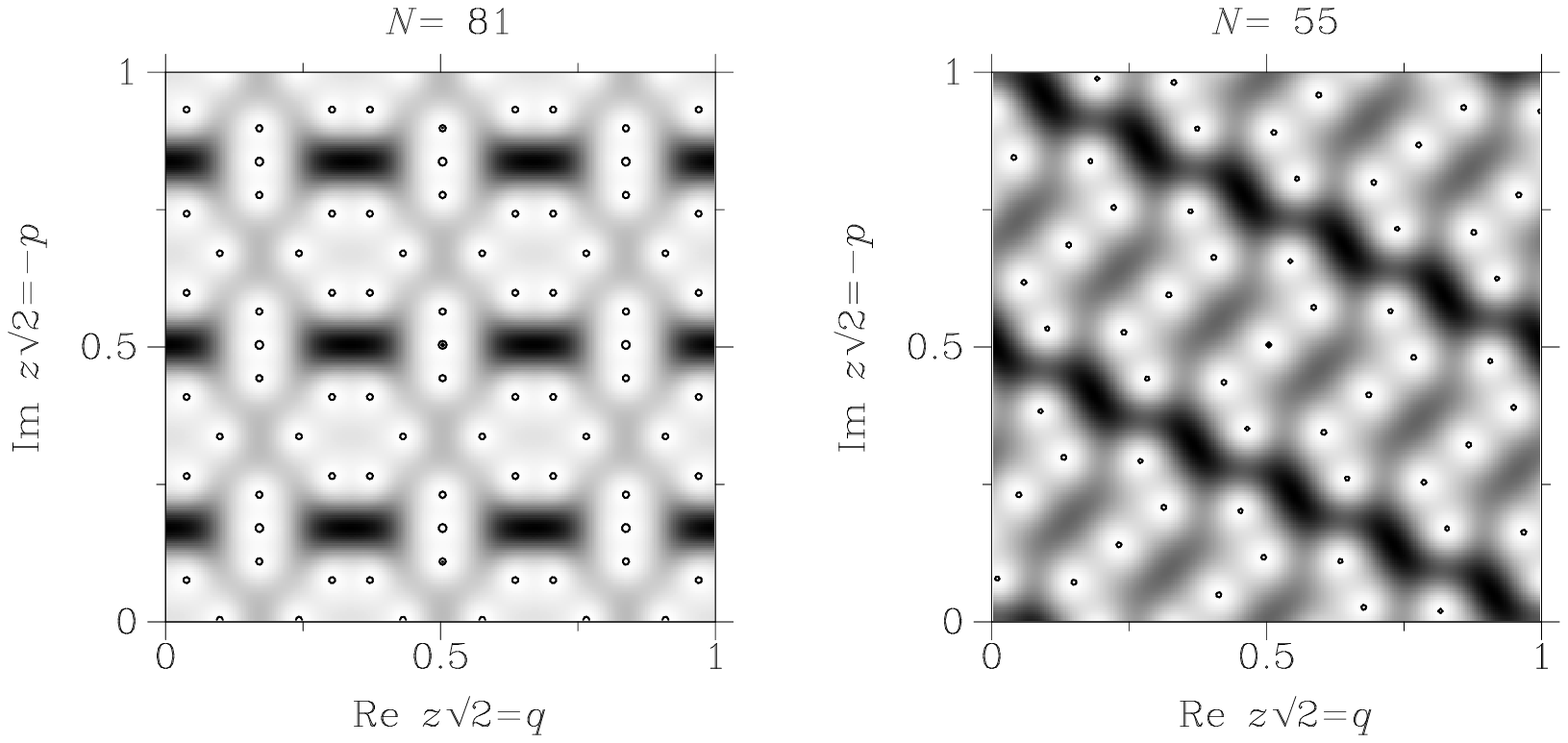}
\caption{}
\end{figure}
\vfill\eject
\begin{figure}
\epsfxsize 17 truecm
\epsfbox{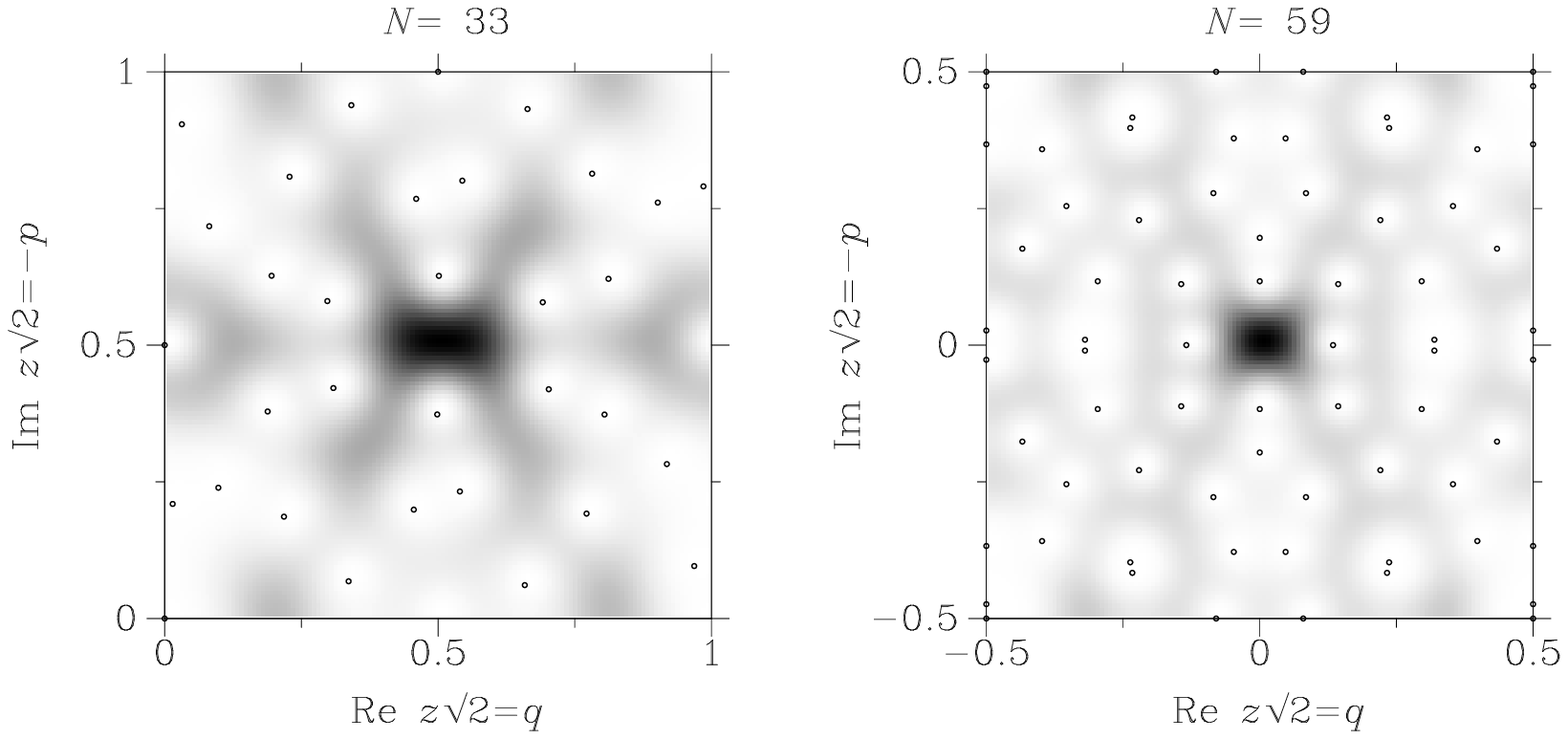}
\caption{}
\end{figure}

\end{document}